\documentclass[acmlarge]{acmart}

\AtBeginDocument{%
}

\usepackage{amsmath}

\usepackage{amssymb}
\usepackage{pifont} 
\usepackage{wasysym} 

\usepackage{graphicx}
\usepackage{xcolor}
\usepackage{colortbl}

\usepackage{tabularx} 
\usepackage{booktabs}
\usepackage{makecell}
\usepackage{multirow}
\usepackage{microtype} 

\usepackage{hyperref}
\hypersetup{
 colorlinks = true,
}
\usepackage[capitalise]{cleveref}

\renewcommand\footnotetextcopyrightpermission[1]{} 

\begin{document}

\title{Collaborative Inference and Learning between Edge SLMs and Cloud LLMs: A Survey of Algorithms, Execution, and Open Challenges}


\author{Senyao Li}
\orcid{0009-0002-4175-4573}
\affiliation{
 \institution{School of Computer Science and Technology, Huazhong University of Science and Technology}
 \city{Wuhan}
 \country{China}
}

\author{Haozhao Wang}
\affiliation{
 \institution{School of Computer Science and Technology, Huazhong University of Science and Technology}
 \city{Wuhan}
 \country{China}
}

\author{Wenchao Xu}
\affiliation{
 \institution{Division of Integrative Systems and Design, Hong Kong University of Science and Technology}
 \city{Hong Kong}
 \country{China}
}

\author{Rui Zhang}
\affiliation{
 \institution{School of Computer Science and Technology, Huazhong University of Science and Technology}
 \city{Wuhan}
 \country{China}
}

\author{Song Guo}
\affiliation{
 \institution{Department of Computer Science and Engineering, Hong Kong University of Science and Technology}
 \city{Hong Kong}
 \country{China}
}

\author{Jingling Yuan}
\orcid{0000-0001-7924-8620}
\email{yjl@whut.edu.cn}
\affiliation{
 \institution{Hubei Key Laboratory of Transportation Internet of Things, Wuhan University of Technology}
 \city{Wuhan}
 \postcode{430070}
 \country{China}
}

\author{Xian Zhong}
\orcid{0000-0002-5242-0467}
\email{zhongx@whut.edu.cn}
\affiliation{
 \institution{Hubei Key Laboratory of Transportation Internet of Things, Wuhan University of Technology}
 \city{Wuhan}
 \postcode{430070}
 \country{China}
}

\author{Tianwei Zhang}
\affiliation{
 \institution{College of Computing and Data Science, Nanyang Technological University}
 \country{Singapore}
}

\author{Ruixuan Li}
\email{rxli@hust.edu.cn}
\affiliation{
 \institution{School of Computer Science and Technology, Huazhong University of Science and Technology}
 \city{Wuhan}
 \country{China}
}


\begin{abstract}
As large language models (LLMs) evolve, deploying them solely in the cloud or compressing them for edge devices has become inadequate due to concerns about latency, privacy, cost, and personalization. This survey explores a collaborative paradigm in which cloud-based LLMs and edge-deployed small language models (SLMs) cooperate across both inference and training. We present a unified taxonomy of edge-cloud collaboration strategies. For inference, we categorize approaches into task assignment, task division, and mixture-based collaboration at both task and token granularity, encompassing adaptive scheduling, resource-aware offloading, speculative decoding, and modular routing. For training, we review distributed adaptation techniques, including parameter alignment, pruning, bidirectional distillation, and small-model-guided optimization. We further summarize datasets, benchmarks, and deployment cases, and highlight privacy-preserving methods and vertical applications. This survey provides the first systematic foundation for LLM-SLM collaboration, bridging system and algorithm co-design to enable efficient, scalable, and trustworthy edge-cloud intelligence.

\end{abstract}




\maketitle

\section{Introduction}

Large language models (LLMs)~\cite{geminiteam2025geminifamilyhighlycapable,abdin2024phi3technicalreporthighly,openai2024gpt4ocard,anthropic273639283model} have demonstrated remarkable proficiency across a broad spectrum of natural language processing tasks. However, their substantial computational and memory demands make on-device deployment on resource-constrained edge devices prohibitive~\cite{xu2024unleashing,zhang2024large,lin2025pushinglargelanguagemodels}. With the rapid advancement of big data, cloud computing, and edge computing, edge-cloud collaborative intelligence has emerged as a promising paradigm for unlocking data value and delivering ubiquitous AI~\cite{niu2025collaborativelearningondevicesmall,pan2024cloud,10488475,hu2024cloud}. This paradigm combines the cloud’s computational power and generalization ability with the edge’s responsiveness and adaptability~\cite{11044698,feng2023knowledge}. 

Rather than merely compressing large models for edge deployment, we focus on the systematic co-optimization of architecture, algorithms, execution, and privacy across cloud-based LLMs and edge-deployed small language models (SLMs)~\cite{llmtune,gu2024minillm,qwen2025qwen25technicalreport,mehta2024openelmefficientlanguagemodel}. By leveraging feature sharing, task partitioning, and knowledge transfer, LLMs and SLMs collaborate to deliver efficient and reliable intelligent services in heterogeneous environments~\cite{10.1145/3636534.3649379,labrak-etal-2024-biomistral,10.5555/3666122.3669417}. Existing cloud-centric methods exploit centralized LLMs for strong generalization but incur high latency, privacy risks, and poor adaptation to localized contexts~\cite{liugrey,cheng2024remoteragprivacypreservingllmcloud,10707974}. In contrast, edge-centric approaches offer fast response and personalization but are limited by on-device compute and generality~\cite{du2022glmgenerallanguagemodel,taylor2022galacticalargelanguagemodel,gemmateam2024gemma2improvingopen,10759588,xu2024unleashing}. Prior surveys have examined cloud-centric, edge-centric, and device-to-device cooperative paradigms, yet each faces key limitations in real-world deployment~\cite{chen2025surveycollaborativemechanismslarge}. The LLM-SLM collaborative paradigm unifies these strengths by enabling hierarchical cooperation between large cloud models and small on-device models~\cite{hao2024hybrid,10764574}.

Unlike federated learning~\cite{10856857,s24248028} or model compression~\cite{zhu2024surveymodelcompressionlarge}, this approach treats LLMs and SLMs as distinct but interactive agents~\cite{liu2025edgecloudcollaborativecomputingdistributed,chen2023eellm}, enabling more dynamic and modular collaboration across the edge-cloud continuum. However, realizing this vision poses several challenges across both inference and training. From the inference perspective, architectural heterogeneity between LLMs and SLMs complicates unified scheduling and deployment; varying task granularities, resource budgets, and latency constraints make coordinated model execution difficult~\cite{chan2024chateval,gur2024a}; and network instability limits high-frequency interactions between models~\cite{li2025pushinglimitmemorybandwidth}, such as model swapping~\cite{jia2025scalingondevicellmsactiveweight} or cross-attention fusion, which are essential for maintaining semantic consistency and responsiveness. On the training side, heterogeneity in data distributions~\cite{liu2024understandingllmscomprehensiveoverview}, task formulations, and model architectures hinders effective knowledge transfer, particularly for distillation and adaptation methods that assume structural or objective alignment. Moreover, the prevalence of non-IID edge data~\cite{s24248028,he2020fedmlresearchlibrarybenchmark} and the need for personalized model behavior exacerbate this issue, as local fine-tuning may lead to overfitting, while centralized updates risk diluting critical edge-specific or causally relevant patterns~\cite{10978177,blendingcrayon}. These intertwined factors collectively underscore the need for more principled designs of collaborative learning and inference frameworks between heterogeneous models.

These limitations motivate a dedicated LLM-SLM collaboration framework that is structurally aware, distributionally robust, and capable of delivering efficient, reliable, and generalizable intelligence under dynamic, resource-constrained conditions.
\begin{figure}
	\centering
	\includegraphics[width = \linewidth]{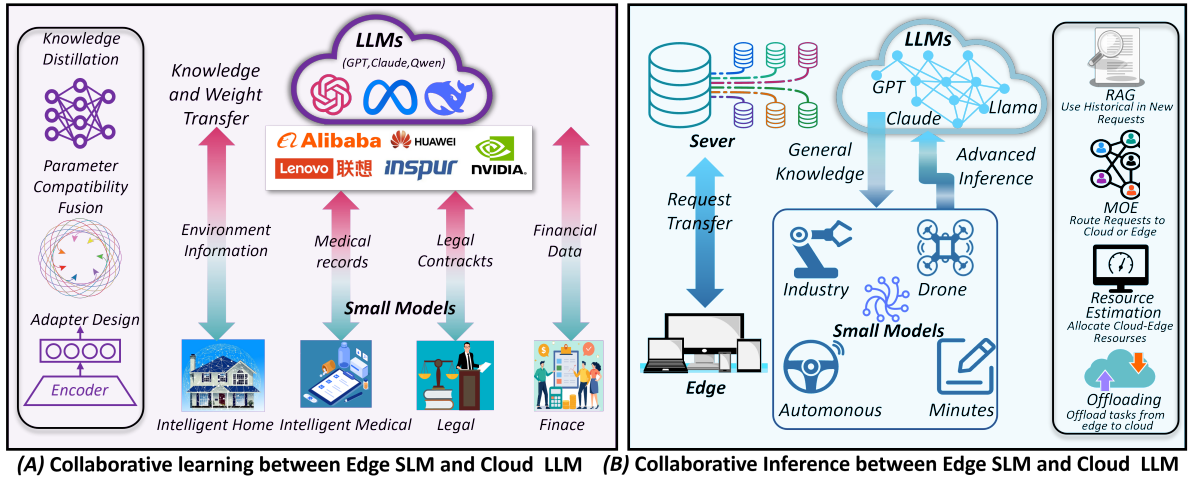}
	\caption{Overview of cloud-edge collaboration workflows: (a) training and (b) inference.}
	\label{fig:1-1}
	\vspace{-5pt}
\end{figure}

In 2021, Alibaba DAMO Academy and Zhejiang University conducted the first large-scale survey on cloud-edge collaboration~\cite{10.1109/TKDE.2022.3178211}, identifying three synergy paradigms. In 2022, Alibaba’s Top Ten Technology Trends highlighted ``co-evolution between cloud-based large models and edge-side small models'', emphasizing intelligent, privacy-aware end-device services~\cite{alibaba2022trends}. Qualcomm echoed this in 2023 with Hybrid AI white papers~\cite{qualcomm2024hybrid,qualcomm2024hybrid2}, advocating intelligent workload partitioning between LLMs and SLMs to address latency, efficiency, privacy, and personalization. These developments reflect a shift from centralized LLM pipelines to hybrid edge-cloud architectures, supported by emerging industrial-scale frameworks. Walle~\cite{lv2022walle} offers an end-to-end system for development, deployment, and runtime, enabling cloud-edge collaboration across 300+ tasks with over 10 billion daily invocations, powering personalized recommendations, multi-modal understanding, and real-time 3D rendering via a lightweight edge inference engine. Luoxi~\cite{luoxi2024models} adopts a ``slow-fast'' learning strategy, where cloud LLMs generate latent representations to assist fast, personalized inference on edge devices with real-time feedback. InfiGUIAgent~\cite{liu2025infiguiagentmulti-modalgeneralistgui} demonstrates hierarchical edge reasoning through two-stage fine-tuning for GUI interaction in multi-modal scenarios. Real-world applications further underscore this paradigm’s necessity: in vision and graphics (urban video analytics~\cite{10415802}, autonomous driving~\cite{hu2024cloud,chen2024edge}, XR), in language tasks (virtual assistants~\cite{openai2024gpt4technicalreport,glm2024chatglmfamilylargelanguage}, input methods, dialogue agents), and in personalized recommendation~\cite{lin2024efficient,lv2025collaboration,long2024diffusion}. Edge intelligence is also expanding into multi-modal and sensor-based domains such as medical diagnosis~\cite{labrak-etal-2024-biomistral}, smart homes~\cite{10621342}, and industrial monitoring, where energy efficiency and responsiveness are critical. Together, these academic and industrial efforts highlight both the technical feasibility and practical urgency of cloud-edge LLM-SLM collaboration, motivating a comprehensive survey of existing paradigms, system designs, and open challenges.

\subsection{Related Surveys and Their Scope}

Xu \textit{et al.}~\cite{xu2024unleashing} provide the first systematic overview of the full AIGC service lifecycle, from data collection and training to inference, and propose a collaborative cloud-edge-end infrastructure for mobile networks. Qu \textit{et al.}~\cite{qu2025mobile} review LLM deployment at the mobile network edge, emphasizing resource-efficient techniques, architectural design, and edge LLM caching. In contrast, Xu \textit{et al.}~\cite{11039635} prioritize system-level optimization, such as model compactification and token pruning, to accelerate inference and enhance deployment efficiency. For task offloading and resource allocation, Wang \textit{et al.}~\cite{10.1145/3284387} model task migration in mobile-cloud offloading, while \cite{10336879} apply reinforcement learning to scheduling in CETCN scenarios. Xu \textit{et al.}~\cite{10398474} survey decentralized continual learning on distributed devices, categorizing three algorithmic strategies for mitigating catastrophic forgetting and distributional shift. Building on this, Niu \textit{et al.}~\cite{niu2025collaborativelearningondevicesmall} comprehensively review collaborative learning mechanisms between device-side small models and cloud-based large models from system, algorithmic, and application perspectives. Under constrained communication, Zhang \textit{et al.}~\cite{zhang2024large} propose air-ground collaborative inference strategies, and Lin \textit{et al.}~\cite{lin2025pushinglargelanguagemodels} introduce a multi-agent LLM framework for natural language tasks in 6G networks. Chen \textit{et al.}~\cite{chen2025surveycollaborativemechanismslarge} systematically summarize LLM-SLM collaboration paradigms, pipelining, routing, distillation, and fusion, as foundational strategies for efficient, adaptive intelligent systems. Existing surveys tend to focus narrowly on system deployment or isolated techniques (\textit{e.g.}, distillation, pipelining) and lack a unified algorithmic perspective on collaborative large-small model training and inference. Crucially, none abstract collaboration into a structured design space or offer a taxonomy capturing functional roles and interaction boundaries between heterogeneous models in edge-cloud settings. 

\begin{figure}
	\centering
	\includegraphics[width = \linewidth]{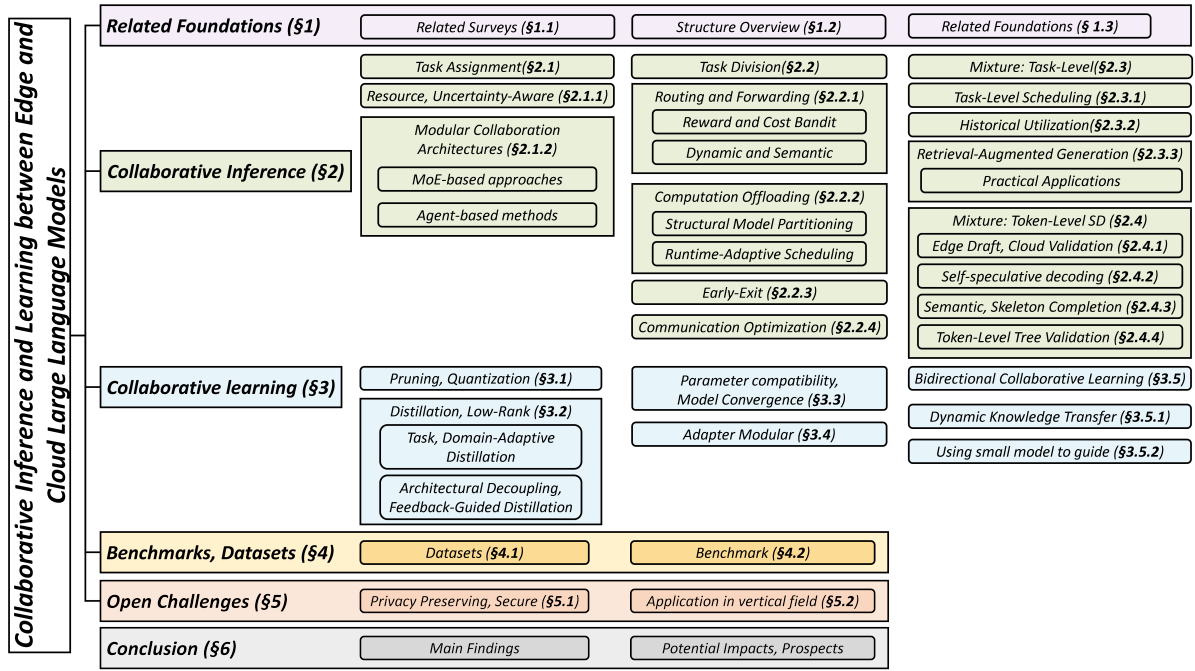}
	\caption{Overview of collaborative paradigms between edge-cloud large and SLMs, structured along two axes: inference cooperation and training collaboration.}
	\label{fig:1-2}
	\vspace{-5pt}
\end{figure}

\begin{table}
	\centering
	\small
	\caption{Comparison of this survey with related reviews}
	\begin{tabularx}{\textwidth}{l|X|X|X}
	\toprule
	\textbf{Reference} & \textbf{Scope} & \textbf{Focus Area} & \textbf{Distinguishing Features} \\
	\midrule
	\multirow{3}{*}{Xu \textit{et al.}~\cite{xu2024unleashing}} 
	& AIGC service lifecycle 
	& Mobile network infrastructure design 
	& Proposes end-cloud-edge architecture; omits model-level inference collaboration. \\
	\midrule
	\multirow{3}{*}{Xu \textit{et al.}~\cite{11039635}} 
	& LLM system deployment 
	& Inference optimization (compression, pruning) 
	& Emphasizes hardware/system acceleration; ignores LLM-SLM interaction logic. \\
	\midrule
	\multirow{3}{*}{Wang \textit{et al.}~\cite{10.1145/3284387,10336879}} 
	& Mobile-cloud task offloading 
	& Scheduling and resource allocation 
	& Models offloading strategies via RL; does not address collaborative inference. \\
	\midrule
	\multirow{3}{*}{Xu \textit{et al.}~\cite{10398474}} 
	& Decentralized continual learning 
	& Distributed algorithmic adaptation 
	& Surveys lifelong learning; lacks explicit LLM-SLM collaboration structure. \\
	\midrule
	\multirow{3}{*}{Niu \textit{et al.}~\cite{niu2025collaborativelearningondevicesmall}} 
	& LLM-SLM collaborative training 
	& System, algorithm, application layers 
	& Reviews collaborative training; does not cover inference across modalities/tasks. \\
	\midrule
	\multirow{4}{*}{Zhang \textit{et al.}~\cite{zhang2024large,lin2025pushinglargelanguagemodels}} 
	& LLMs in constrained networks (6G) 
	& Multi-agent and aerial-ground collaboration 
	& Focuses on special network conditions; proposes communication-specific inference strategies. \\
	\midrule
	\multirow{3}{*}{Chen \textit{et al.}~\cite{chen2025surveycollaborativemechanismslarge}} 
	& LLM-SLM collaboration mechanisms 
	& Pipelining, routing, distillation, fusion 
	& Summarizes interaction methods; lacks a unified algorithmic design space for inference. \\
	\midrule
	\rowcolor{gray!20}
	\multirow{5}{*}{This Survey} 
	& Heterogeneous LLM-SLM collaboration 
	& Inference and training algorithms 
	& First to cover end-cloud collaborative inference (task assignment/division/mixture) and training (parameter fusion, knowledge transfer). \\
	\bottomrule
	\end{tabularx}
	\label{tab:survey_comparison}
\end{table}

\textbf{In contrast, this survey presents the first comprehensive review that jointly considers inference-time collaboration and training-time coordination between LLMs and SLMs.} As shown in \cref{fig:1-1} and \cref{tab:survey_comparison}, we propose a unified taxonomy of collaboration paradigms, encompassing task assignment, task division, and mixture (with task-level and token-level granularities), and analyze their alignment with algorithmic principles and system constraints. On the training side, we summarize collaborative adaptation methods, including bidirectional distillation, quantization, pruning, and low-rank approximation, which facilitate efficient SLM deployment without sacrificing performance. This survey places particular emphasis on the interaction between LLMs and SLMs across edge and cloud environments. By jointly analyzing collaboration across both inference and training phases, our review bridges algorithmic design and deployment needs, offering methodological insights and practical implications for future edge-cloud LLM systems. Moreover, we ground our discussion in recent literature published since 2023, ensuring both coverage and relevance. By jointly addressing both inference and training dimensions, our survey fills a key methodological gap and establishes a generalizable framework for designing large-small model cooperation in heterogeneous, resource-constrained environments.

\subsection{Research Questions and Survey Structure}

This survey systematically reviews the development, challenges, and future directions of edge-cloud collaborative inference and training with large and SLMs. We organize the discussion around the following core research questions (RQs):

\begin{itemize}
	\item \textbf{RQ1} \textit{What is the edge-cloud collaborative inference paradigm with large and small models? What are its fundamental concepts and system architecture?} 
	We focus on the basic paradigm, collaboration roles, and system-level design.

	\item \textbf{RQ2} \textit{What are the major paradigms and collaboration patterns in edge-cloud inference?} 
	We cover task scheduling, mixture cooperation at task and token levels, and speculative decoding. Examines how collaboration reduces latency, adapts to uncertainty, and improves efficiency.

	\item \textbf{RQ3} \textit{Why and how should we study collaborative training between large and small models in edge-cloud environments?} 
	We explore the necessity of training collaboration under heterogeneity, summarizing paradigms such as distillation, modular tuning, and parameter alignment for cross-device adaptation.

	\item \textbf{RQ4} \textit{Why is it important to review benchmarks, privacy-preserving methods, and vertical applications in the context of edge-cloud collaboration?} 
	We address fair evaluation standards, privacy challenges in training and inference, and practical requirements across real-world domains.
 
\end{itemize}

As shown in \cref{fig:1-2}, we structure the survey as follows:

\begin{itemize}
	\item \textbf{\cref{Related Concepts and Foundations}} introduces the fundamentals of edge-cloud collaboration with large and SLMs, covering system architectures, edge device constraints, and trends in large-model development (addresses RQ1).

	\item \textbf{\cref{Overview of Collaborative Inference}} examines task assignment and inference strategies, with emphasis on dynamic scheduling and draft-refine verification frameworks, illustrating how edge and cloud models execute inference collaboratively and efficiently (addresses RQ2).

	\item \textbf{\cref{Collaborative-learning-Architectures}} surveys cloud-edge training architectures and deployment strategies, including quantization, pruning, parameter-efficient fine-tuning, and bidirectional knowledge transfer, to support efficient SLM deployment without sacrificing performance (addresses RQ3).

	\item \textbf{\cref{Benchmarks-Datasets-and-Evaluation-Protocols} - \cref{Open Challenges}} summarize benchmarks, evaluation metrics, and domain-specific deployments to facilitate model comparison and practical implementation, and highlight open challenges such as generalizability, privacy, sustainability, and multi-agent collaboration (addresses RQ4).

	\item \textbf{\cref{Conclusion}} concludes with key insights and future research directions.
 
\end{itemize}

\subsection{Related Concepts and Foundations}
\label{Related Concepts and Foundations}

\subsubsection{Definition and Characteristics of LLMs}

LLMs typically refer to transformer-based architectures with billions to trillions of parameters. These models are pretrained on massive-scale corpora using strategies such as autoregressive learning (\textit{e.g.}, GPT~\cite{openai2024gpt4technicalreport,openai2024gpt4ocard}) or masked language modeling (\textit{e.g.}, Qwen~\cite{qwen2025qwen25technicalreport}, DeepSeek~\cite{deepseekai2025deepseekv3technicalreport}). Architecturally, LLMs stack tens to hundreds of transformer blocks, each comprising multi-head attention, feed-forward layers, residual connections, and normalization. As scale grows, LLMs exhibit emergent abilities, such as symbolic reasoning, instruction following, and multimodal understanding, that do not scale linearly with parameter count. In edge-cloud collaboration, LLMs are typically cloud-hosted due to their intensive computational and memory needs. While offering strong generalization and zero-/few-shot transfer across tasks, they also introduce significant communication overhead, as each inference often requires multiple cloud interactions, increasing latency and bandwidth load. Additionally, their training and fine-tuning involve high I/O costs and poor device adaptability, limiting their use in real-time or privacy-sensitive edge scenarios.

\subsubsection{Definition and Characteristics of SLMs}

SLMs, typically comprising millions to a few hundred million parameters, are built for resource-constrained environments while retaining core language modeling abilities. They simplify architecture via shallower transformer stacks, smaller hidden sizes, and fewer attention heads, greatly reducing memory and power use, enabling efficient deployment on mobile devices, microcontrollers, and edge nodes. SLMs emphasize responsiveness and local availability over broad reasoning. Though they lack LLMs’ emergent abilities, they perform well in context-specific tasks like smart input, offline assistants, and embedded systems. Through knowledge distillation, LoRA fine-tuning, and instruction tuning, SLMs can approximate LLM outputs in narrow domains while maintaining low latency and data locality. In collaborative setups, they act as first-response agents or draft generators, offloading complex reasoning to cloud LLMs only when needed, thus optimizing communication cost and model efficiency.

\subsubsection{Complementarity of hardware architectures and application scenarios}

The foundation for cloud-edge LLM-SLM collaboration stems from hardware-driven capability partitioning: Cloud servers leverage multi-GPU clusters (\textit{e.g.}, NVIDIA H100) with high-bandwidth interconnects (NVLink/RoCE) to run trillion-parameter models (\textit{e.g.}, LLaMA-3 405B) for complex reasoning and global data processing~\cite{DBLP:conf/hpca/SunGLZCZWL25}, while resource-constrained edge devices, spanning smartphones (6-16GB RAM), XR headsets, and embedded systems (\textit{e.g.}, Jetson Orin), deploy billion-parameter SLMs (\textit{e.g.}, TinyBERT~\cite{DBLP:conf/emnlp/JiaoYSJCL0L20}) for latency-critical tasks (<10ms response) and local data optimization~\cite{DBLP:conf/hpca/SeoK0YMPL25}. This architectural complementarity enables hierarchical task allocation: SLMs execute real-time perception (keyword detection, speech-to-text), privacy-sensitive preprocessing (medical data anonymization), and intent filtering (90\% accuracy in customer service), offloading complex generation/reasoning to LLMs. Technical synergies like knowledge distillation (TinyBERT achieving 96\% of BERT’s performance~\cite{DBLP:conf/emnlp/JiaoYSJCL0L20}) and parameter-efficient tuning (LoRA boosting medical QA accuracy by 20\%) further bridge capability gaps, with federated learning ensuring privacy when SLMs transmit anonymized features to cloud LLMs for multimodal fusion, collectively enabling 40\% efficiency gains in applications like in-vehicle voice systems (200ms latency) and healthcare diagnostics.

\begin{table}
	\small
	\centering
	\caption{Comparison of major paradigms in cloud-edge collaborative inference}
	\begin{tabularx}{\linewidth}{l|X|X|X}
	\toprule
	\multirow{2}{*}{\textbf{Category}} & \textbf{Definition \& Characteristics} & \multirow{2}{*}{\textbf{Advantages}} & \multirow{2}{*}{\textbf{Limitations}} \\
	\midrule
	\multirow{3}{*}{\textbf{Task assignment~\S~\ref{sec:task-assignment}}} 
	& Routes requests to SLM or LLM based on confidence, resource, or task type. 
	& Simple and fast; minimal communication overhead. 
	& Hard switching may misroute uncertain inputs; lacks joint reasoning. \\
	\midrule
	\multirow{3}{*}{\textbf{Task division~\S~\ref{sec:Task Division}}} 
	& Splits tasks into modules (\textit{e.g.}, early exit, routing) processed by SLM and LLM. 
	& Supports modular, adaptive inference with fine-grained control. 
	& Requires explicit segmentation; incurs coordination overhead. \\
	\midrule
	\multirow{3}{*}{\textbf{Mixture: Task-level~\S~\ref{sec:mixture-task}}} 
	& Combines assignment and division across stages or roles via communication. 
	& Flexible decomposition; leverages complementary strengths. 
	& Dependent on partitioning quality and coordination efficacy. \\
	\midrule
	\multirow{3}{*}{\textbf{Mixture: Token-level~\S~\ref{sec:mixture-token}}} 
	& Collaborates at token generation (\textit{e.g.}, speculative decoding). 
	& Low-latency with accurate output; efficient cloud fallback. 
	& Sensitive to draft quality; complex fusion and validation. \\
	\bottomrule
	\end{tabularx}
	\label{tab:Major Paradigms}
\end{table}

\section{Overview of Collaborative Inference}
\label{Overview of Collaborative Inference}

Collaborative inference between edge-deployed SLMs and cloud-based LLMs seeks to optimize latency and service efficiency under bandwidth and responsiveness constraints~\cite{chen2025harnessingmultiplelargelanguage}. As shown in \cref{tab:Major Paradigms}, recent work has explored three complementary collaboration paradigms. It is important to note that the classification presented here focuses specifically on inference-time collaboration between heterogeneous models. Although some mechanisms, such as mixture-of-experts (MoE), are applicable to both inference and training phases, this survey confines its scope to inference-oriented scenarios. That is, we summarize and compare collaborative strategies that are either designed for or primarily evaluated under inference workloads. We categorize and review existing works from the following perspectives:
\begin{itemize}
	\item \textbf{Task assignment}, which routes queries to the most suitable model (\textit{e.g.}, via agent-based or MoE strategies);
	\item \textbf{Task division}, which decomposes execution across models or system components (\textit{e.g.}, offloading, request routing, or early-exit mechanisms);
	\item \textbf{Mixture strategies}, which integrate assignment and division at task-level or token-level granularity using techniques such as speculative decoding and multi-stage validation.
\end{itemize}

\begin{figure}
	\centering
	\includegraphics[width = \linewidth]{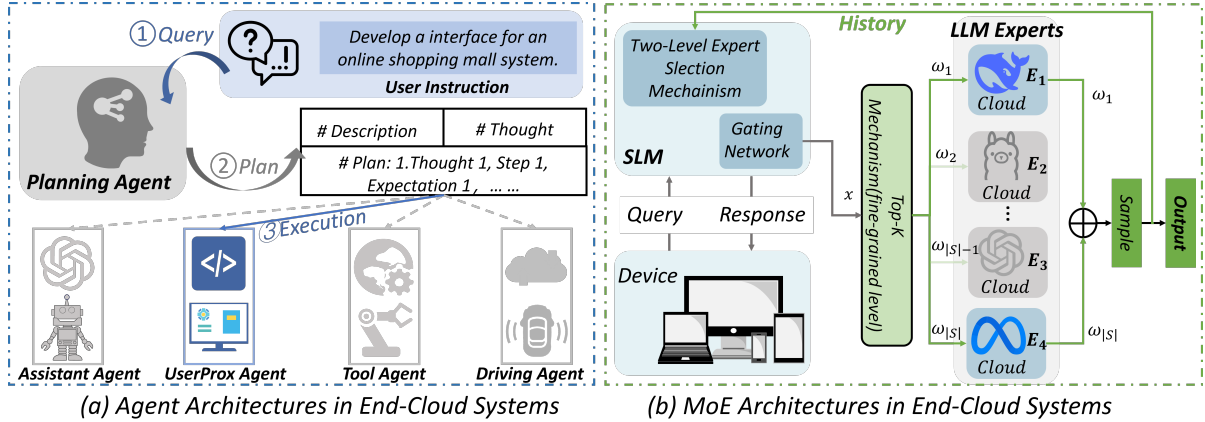}
	\caption{Comparison of task assignment in edge-cloud systems: (a) Agents decompose user instructions to multi-stage plans executed by specialized agents; (b) MoE-based selection and fusion of top-$k$ cloud LLM experts via fine-grained gating.}
	\label{fig:agent-MoE}%
	\vspace{-5pt}
\end{figure}

\subsection{Task Assignment: Dynamic Model Selection for Cost-Quality Trade-offs}
\label{sec:task-assignment}

Task assignment determines whether a request is processed entirely by an edge-side SLM or a cloud-based LLM, in contrast to collaborative decoding, which involves inter-model interaction. To minimize latency and energy consumption under quality-of-service (QoS) constraints, prior methods employ lightweight scorers, calibrated reward estimators, or bandit-based controllers to select the optimal execution path before generation. Some approaches further incorporate user preference vectors or learned model-capability representations for adaptive, runtime routing. FS-GEN~\cite{zhang2024fastslow} introduces \textit{collaboration frequency} to quantify the trade-off between large and small models in generation tasks: by aligning input spaces and defining a unified output-fusion objective, FS-GEN measures how often the cloud LLM intervenes in frequency-dominated sequences. Broadly, task assignment strategies fall into three architectural paradigms, resource- and uncertainty-aware assignment, agent-based scheduling and mixture-of-experts (MoE) frameworks (see \cref{fig:agent-MoE}), each reflecting different coordination granularities and model-selection philosophies.

\begin{table}
	\small
	\centering
	\caption{Resource- and uncertainty-aware task assignment strategies}
	\begin{tabularx}{\linewidth}{l|X|X|X}
	\toprule
	\textbf{Reference} & \textbf{Key Idea} & \textbf{Advantage} & \textbf{Limitation} \\
	\midrule
	\multirow{2}{*}{PerLLM~\cite{yang2024perllm}} 
	& Personalized scheduling with constraint satisfaction 
	& Jointly optimizes deployment and resource usage 
	& Complex constraint modeling \\
	\midrule
	\multirow{2}{*}{EdgeLLM~\cite{cai2024edge}} 
	& Value-density-first scheduling with preemption and batching 
	& Cost-effective, real-time task prioritization 
	& Preemption can destabilize under dynamic loads \\
	\midrule
	\multirow{3}{*}{FS-GEN~\cite{zhang2024fast}} 
	& Dual-system inference with uncertainty-triggered LLM fallback 
	& Latency-efficient by conditional large-model invocation 
	& Accuracy depends on uncertainty-estimation precision \\
	\midrule
	\multirow{2}{*}{Yang \textit{et al.}~\cite{yang2024efficient}} 
	& Operator-level scheduling based on compute intensity 
	& Fine-grained control of module placement 
	& Requires precise operator profiling \\
	\midrule
	\multirow{2}{*}{Stammler \textit{et al.}~\cite{10387895}} 
	& Hybrid cost-aware branching with fault-tolerant rollback 
	& Reduces re-execution cost during cloud failures 
	& Overhead in maintaining state consistency \\
	\midrule
	\multirow{2}{*}{U-VPA~\cite{gan2023cloud}} 
	& Uncertainty-guided sampling in a teacher-student setup 
	& Bandwidth-efficient selection of informative samples 
	& Needs robust domain-shift detection \\
	\midrule
	\multirow{2}{*}{Li \textit{et al.}~\cite{li2018auto,10621342}} 
	& INT8 model partitioning with automatic tuning 
	& Efficient, low-latency dynamic deployment 
	& Precision loss under aggressive quantization \\
	\midrule
	\multirow{2}{*}{KDSL~\cite{chen2025knowledgede}} 
	& LLM-based rule generation with feedback verification 
	& Closed-loop correction with minimal cloud calls 
	& Complexity in evaluating rule quality \\
	\bottomrule
	\end{tabularx}
	\label{tab:task Allocation}
\end{table}

\subsubsection{Resource- and Uncertainty-Aware Task Assignment}

To leverage heterogeneous models while respecting edge-cloud capacity constraints (see \cref{tab:task Allocation}), recent work explores diverse task allocation mechanisms that handle runtime uncertainty, system heterogeneity, and model capacity:
FS-GEN~\cite{zhang2024fast} adopts a dual-system architecture where fast but uncertain SLMs serve as \textit{System 1}, and reliable LLMs act as \textit{System 2}, invoked when ambiguity arises. Fang \textit{et al.}~\cite{fang2023large} propose a feedback-driven framework that adjusts routing decisions dynamically based on runtime uncertainty, model confidence, and system state, avoiding static thresholds.
EdgeLLM~\cite{cai2024edge} uses a value-density-first algorithm to rank tasks by cost-effectiveness, supporting preemption and batched execution via adaptive thresholds. Yang \textit{et al.}~\cite{yang2024efficient} introduce fine-grained module-level partitioning, offloading heavy components (\textit{e.g.}, attention, linear layers) to the cloud while retaining lightweight operations at the edge.
U-VPA~\cite{gan2023cloud} employs uncertainty-guided sampling within a teacher-student framework, selectively uploading domain-informative samples to enhance generalization under non-stationary distributions.
A hybrid cost function~\cite{10387895} identifies branch points for edge-cloud transitions, with fault-tolerant re-execution of failed branches locally. Ye \textit{et al.}~\cite{10621342} dynamically select intermediate INT8 representations for edge deployment via automatic tuning.
KDSL~\cite{chen2025knowledgede} leverages LLMs to generate logical rules via UCT-guided search. These rules are verified and deployed on the edge, enabling symbolic reasoning to refine cloud predictions.

\subsubsection{Modular Collaboration via MoE and Agent Architectures}

\paragraph{MoE-based approaches.} 
Mixture-of-Experts (MoE) architectures enable scalable, adaptive inference in edge-cloud settings. A lightweight on-device SLM routes queries via a gating or routing network to activate a sparse set of cloud LLM experts (see \cref{fig:agent-MoE}), whose responses are fused for soft assignment and diversity. ${\mathrm{MoE}}^2$~\cite{jin2025moe} and EdgeMoE~\cite{yi2023edgemoe} guide expert routing using statistical priors or learned policies to reduce memory and I/O overhead. LiteMoE~\cite{10.1145/3666025.3699355} further reduces cost by identifying and merging critical experts without retraining. Extending selection to structuring, DoT~\cite{shao2025division} uses a task decomposition module and dependency-graph scheduler to partition complex queries into subtasks, enabling adaptive granularity and critical-path prioritization. Tian \textit{et al.}~\cite{tian-etal-2024-dialogue} propose role-aware MoE routing for multi-turn dialogue, assigning segments to specialized experts and fusing outputs for contextually rich generation. CoEL~\cite{li2025moeempowerededgellmsdeployment} adds resource-adaptive coordination across edge devices, supporting elastic deployment and efficient scaling under varying budgets.

\paragraph{Agent-based methods.} 
These methods use modular execution pipelines in which a planning agent interprets instructions, formulates structured plans, and delegates subtasks to specialized agents~\cite{wang2025comprehensivesurveyllmagentstack} (\textit{e.g.}, assistants, tools, drivers) for context-aware, goal-driven coordination~\cite{guo2024algorithmicscomplexitycostdriventask}. ARAG~\cite{maragheh2025aragagenticretrievalaugmented} integrates four agents, user understanding, natural language inference (NLI), context summarization, and item ranking, into a RAG pipeline. AgentVerse~\cite{chen2024agentverse} deploys an edge-based coordinator to assemble cloud expert teams by task intent, supporting horizontal aggregation (\textit{e.g.}, voting) and vertical communication for complex reasoning. Salve \textit{et al.}~\cite{salve2024collaborativemultiagentapproachretrievalaugmented} propose a specialized multi-agent framework with a central module orchestrating context-aware query generation and multi-source retrieval. In physically grounded scenarios, WebAgent~\cite{gur2024a} fuses LLMs with an embodied VirtualHome agent for goal-directed exploration and physical reasoning. ChatEval~\cite{chan2024chateval} introduces a debate-style framework where LLM-based judges collaboratively assess generation quality. EcoAgent~\cite{yi2025ecoagentefficientedgecloudcollaborative} implements a closed-loop system with a cloud planner and two edge agents, one for execution and one for result verification, that triggers replanning upon failure. Finally, the MADRL framework~\cite{10185648} enables centralized cloud training and decentralized edge execution, reducing training overhead and inference latency in multi-agent reinforcement learning~\cite{10488475}.

\begin{table}
	\small
	\centering
	\caption{Comparison of collaborative inference paradigms: routing, offloading, early exit, and communication optimization}
	\begin{tabularx}{\linewidth}{l|X|X|X}
	\toprule
	\textbf{Paradigm} & \textbf{Core Innovation} & \textbf{Typical Scenario} & \textbf{Limitations} \\
	\midrule
	\multirow{4}{*}{\textbf{{Routing and Forwarding}~\S~\ref{Routing and Forwarding Mechanisms}}}
	& Combines confidence, semantic planning, and contextual bandits for query routing.
	& Budget- or latency-constrained query routing.
	& Depends on accurate confidence estimates; fallback may add overhead or misroute. \\
	\toprule
	\multirow{4}{*}{\textbf{{Computation Offloading}~\S~\ref{Computation Offloading and System-Level Optimization}}}
	& Supports dynamic scheduling and token-/layer-wise partitioning of models.
	& Latency-sensitive tasks under fluctuating resource availability.
	& Performance degrades in highly dynamic or non-stationary environments. \\
	\toprule
	\multirow{4}{*}{\textbf{{Early Exit}~\S~\ref{Early Exit}}}
	& Enables token- or layer-level early termination via dropout and pipeline reuse.
	& Low-latency generation or streaming applications.
	& Accuracy can drop at exit points; modeling token-dependent dynamics is challenging. \\
	\toprule
	\multirow{4}{*}{\textbf{{Communication Optimization}~\S~\ref{Communication Optimization}}}
	& Uses entropy compression, bandit routing, and adaptive switching to reduce transfer.
	& Resource-constrained settings with frequent edge-cloud interactions.
	& Balancing compression and semantic fidelity remains difficult; sensitive to distribution shifts. \\
	\bottomrule
	\end{tabularx}
	\label{tab:division_paradigms}
\end{table}

\subsection{Task Division: Cooperative Subtask Decomposition Between Large and Small Models}
\label{sec:Task Division}

\begin{figure}
	\centering
	\includegraphics[width = 0.9\linewidth]{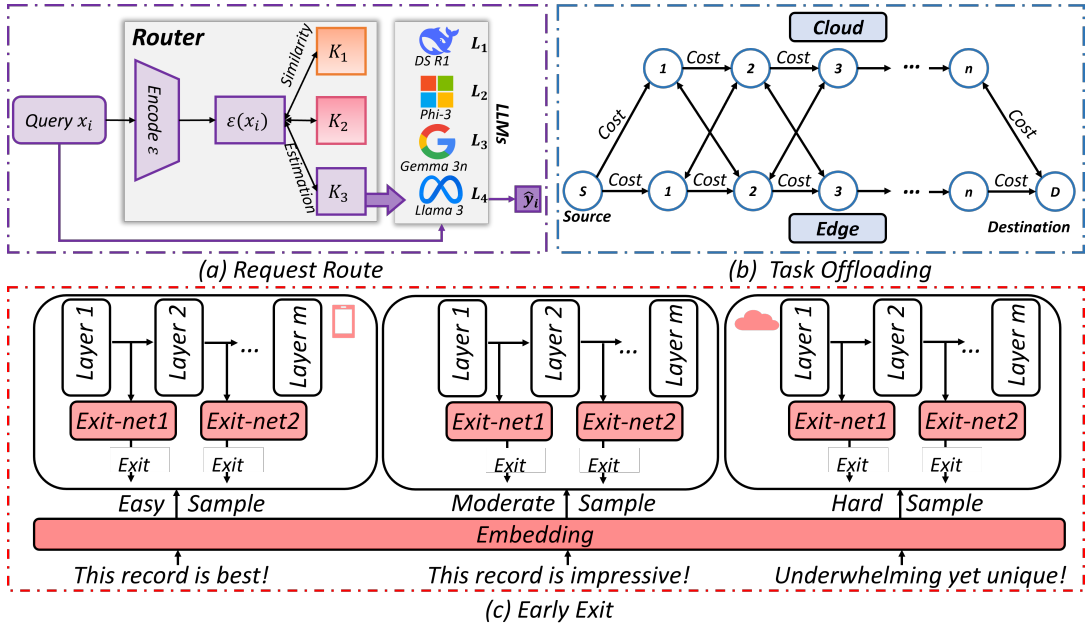}
	\caption{Illustration of three execution-phase task-division strategies: (a) request routing, dispatching queries to optimal LLM experts based on semantics and cost; (b) task offloading, splitting computation graphs across edge and cloud for cost-aware load balancing; and (c) early exit, multi-exit networks that terminate early for simple inputs to reduce latency and energy.}
	\label{fig:Route-off-Exit}
	\vspace{-5pt}
\end{figure}

Task division breaks hierarchical or modular tasks into semantic or functional parts, enabling LLMs and SLMs to jointly execute complementary subtasks. This cooperation improves parallelism and responsiveness under edge constraints. Unlike the categorization~\cite{king2024thoughtfulthingsbuildinghumancentric}, we identify three fundamentally distinct paradigms of task division in edge-cloud collaborative inference, routing, offloading, and early exit, as illustrated in \cref{fig:Route-off-Exit}. Each paradigm offers unique advantages and is suited to different application scenarios, as summarized in \cref{tab:division_paradigms}.

\subsubsection{Routing and Forwarding Mechanisms}
\label{Routing and Forwarding Mechanisms}

Routing-based methods dynamically select models at inference time to balance latency, cost, and accuracy:

\paragraph{Trust- and Semantic-Aware Routing.} 
These approaches use runtime confidence estimation and semantic planning for adaptive model selection. FrugalGPT~\cite{chen2023frugalgptuselargelanguage} reduces API costs via prompt compression and cascaded routing. Tabi~\cite{10.1145/3552326.3587438} employs calibrated confidence scores to choose between lightweight models and powerful LLMs. Dekoninck \textit{et al.}~\cite{dekoninck2025unifiedapproachroutingcascading} integrate pre- and post-inference quality estimation under a unified framework, improving routing accuracy and fallback success. Kag \textit{et al.}~\cite{kag2023efficient,hu2025adaptlink} introduce backup-block preloading for robust failover across heterogeneous devices. RouteLLM~\cite{ong2025routellm} organizes models hierarchically and applies dynamic programming to minimize tiered performance disparity.

\paragraph{Reward- and Cost-Aware Bandit Routing.} 
These methods use online learning to optimize routing based on reward signals and cost-performance trade-offs. HybridLLM~\cite{ding2024hybrid} defines a quality gap and uses a lightweight encoder to predict routing probabilities. ZOOTER~\cite{lu-etal-2024-routing} leverages query-level rewards for expert selection. RouterDC~\cite{NEURIPS2024_7a641b8e} uses dual contrastive loss to align queries with high-performing models. LLM Bandit~\cite{li2025llmbanditcostefficientllm} introduces ``model identity vectors'' for preference-conditioned routing, while MixLLM~\cite{wang-etal-2025-mixllm} and CITER~\cite{zheng2025citercollaborativeinferenceefficient} employ contextual bandits to estimate cost and quality. RouteT2I~\cite{xin2024edgecloudroutingtexttoimagemodel} selects between edge and cloud models for text-to-image tasks using multi-dimensional quality metrics.

\subsubsection{Computation Offloading and System-Level Optimization}
\label{Computation Offloading and System-Level Optimization}

Offloading divides inference tasks across the device and cloud based on runtime conditions: rather than statically assigning tasks from the outset (as in routing), offloading makes dynamic, stage-wise decisions during execution, evaluating current workload against system capacity to maximize performance and utilize resources cost-effectively. We categorize offloading into two complementary types: 

\paragraph{Structural Model Partitioning.} 
This strategy decomposes inference by structurally partitioning models across device and edge/cloud at selected layers or token stages according to computational cost and depth~\cite{DBLP:journals/tcc/YuanXDZM23}. ADAS~\cite{hu2024cloud} leverages internet of things (IoT) networks and proposes an enhanced task offloading algorithm based on DDPG, where a diffusion model is employed to generate noise and determine the optimal execution location for each task (\textit{i.e.}, cloud, edge, or local). Liu \textit{et al.}~\cite{liu2024resource} assign lower layers to local devices and offload higher reasoning to edge servers. CE-CoLLM~\cite{jin2024collm} uses token-level confidence to decide which tokens to process locally. Li \textit{et al.}~\cite{li2018auto} automatically select INT8-quantized intermediate layers as partition points.

\paragraph{Runtime-Adaptive Scheduling.} 
This group of methods dynamically schedules offloading decisions by adapting to real-time metrics, such as latency, confidence, and resource availability, and distributes tasks across local devices, control units, and auxiliary vehicles. Hao \textit{et al.}~\cite{9784600,svirschevski2024specexec} use device metrics and confidence scores for fine-grained control. He \textit{et al.}~\cite{he2024large} introduce a reward-free policy based on latent states. Enhanced Hybrid Inference~\cite{peng2024enhanced} incorporates user-aware utility models under bandwidth constraints. AVA~\cite{10720029} combines federated and multi-agent reinforcement learning for multi-tier distribution. While these approaches excel at adaptive, condition-aware scheduling, they often struggle to generalize in highly dynamic or heterogeneous environments.

\subsubsection{Early Exit}
\label{Early Exit}

Early-exit is a hybrid inference paradigm in which execution can terminate at intermediate layers based on system load or model confidence, reducing computation through dynamic exit decisions. LITE~\cite{varshney-etal-2024-investigating} introduces a confidence-guided early-exit strategy that halts inference without sacrificing generation quality. LayersKip~\cite{elhoushi-etal-2024-layerskip} applies progressively increasing layer dropout and an early-exit loss across Transformer layers to support reliable early termination. EE-LLM~\cite{chen2023eellm}, designed for 3D-parallelized LLMs, integrates token-level exits via key-value recomputation and pipeline parallelism~\cite{10.1145/3698767}, making it compatible with both large-scale training and inference. EESD~\cite{liu2024speculative} uses the initial $N$ layers of a base model and appends a single-layer Transformer to efficiently generate high-quality draft tokens. FREE~\cite{bae-etal-2023-fast} introduces temporally coherent parallelism by synchronizing shallow processing units with precomputed residuals, accelerating token generation with minimal depth. Collectively, these methods reduce latency and cost but still struggle to maintain consistent accuracy across exit points and coordinate exits under token-dependent dynamics.

Efficient communication is critical for collaborative inference in edge-cloud systems, enabling fine-grained reasoning through joint task division and communication-aware scheduling. Hu \textit{et al.}~\cite{hu2025hybrid} propose a hybrid architecture that dynamically delegates inference between lightweight edge models and powerful cloud LLMs, achieving 91\% diagnostic accuracy with a 28.6\% reduction in energy consumption. LLMCascades~\cite{yue2024large} uses a multi-stage framework in which edge predictions are either accepted or escalated to cloud models via voting and verification, enhancing trustworthiness with minimal redundancy. EdgeShard~\cite{10818760} reduces unnecessary cloud queries by forwarding only inference-critical token features, while entropy-based compression~\cite{shen2024large} further minimizes transmission cost without accuracy loss. For throughput optimization, PipeEdge~\cite{9996638} partitions LLMs across devices using pipeline parallelism and automated scheduling, reducing latency while preserving model capacity. PerLLM~\cite{yang2024perllm} frames coordination as a constrained multi-armed bandit problem, employing an upper confidence bound algorithm to adaptively select the most cost-effective execution path. Blending~\cite{lu2024blendingneedcheaperbetter} enables turn-wise model switching in multi-turn interactions, leveraging model heterogeneity to improve response quality while maintaining cost efficiency. Despite these advances, minimizing overhead in dynamic environments and preserving semantic coherence across partitioned reasoning remain open challenges.

\subsubsection{Communication Optimization}
\label{Communication Optimization}

\subsection{Mixture: Task-Level Orchestration and Delegation in LLM-SLM Collaborative Inference}
\label{sec:mixture-task}

\begin{figure}
	\centering
	\includegraphics[width = \linewidth]{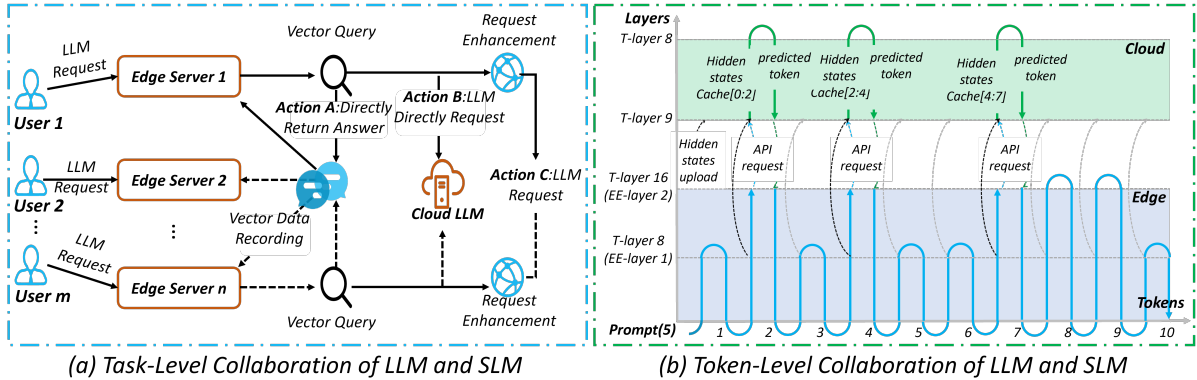}
	\caption{Mixture strategies for edge-cloud model collaboration: (a) task-level, edge servers decide to respond locally, query the cloud, or forward to LLMs based on context and history; (b) token-level, edge SLMs generate easy tokens locally and offload harder ones to cloud LLMs via hidden-state sharing for fine-grained inference.}
	\label{fig:task-token-colab}
	\vspace{-5pt}
\end{figure}

The \textit{mixture paradigm} hybrids task assignment and execution division, allowing SLMs and LLMs to cooperatively handle a request through staged responsibility sharing. As shown in \cref{fig:task-token-colab}, we distinguish two granularities: \textit{1) task-level mixture}, \textit{2) token-level mixture}, where generation is shared step-by-step across models~\cite{chen2024towards,chen2024adaptivelayersplittingwireless,10118601}.

\subsubsection{Task-Level Decomposition, Scheduling, and Orchestration}
\label{sec:Task-Level}

Task-level mixture uses semantic cues to identify subtask boundaries and assign them to edge or cloud models based on computational requirements, yielding interpretable, modular workflows. For example, MinionS~\cite{narayan2025minionscost} pipelines edge-side decomposition with cloud aggregation, while HybridSD~\cite{yan2024hybrid} offloads high-level structural reasoning to the cloud and refines perceptual details on-device. IntellectReq~\cite{lv2024intelligent} generates abstract intents in the cloud for lightweight edge execution. BAIM~\cite{10577141} integrates multiple edge-model outputs via a gating network for multi-modal collaboration, and OT-GAH~\cite{10759588} uses online tree pruning and assignment heuristics to maximize batch throughput. To streamline chain-of-thought reasoning, HAWKEYE~\cite{she2025hawkeyeefficientreasoningmodelcollaboration} applies length-penalized reinforcement learning in the LLM and offloads detailed expansions to the SLM.

\subsubsection{Historical-Enhancement Collaborative Inference}

Users’ queries to cloud APIs often mirror those sent to local models, motivating methods that leverage historical user-cloud interactions to improve on-device inference. Existing approaches include retrieval-augmented generation, cache-based scheduling, and collaborative learning. For instance, SlimPLM~\cite{tan-etal-2024-small} uses heuristic confidence scores to trigger multi-stage retrieval only when the local response is unreliable. VELO~\cite{yao2024velo} caches prior request embeddings on edge servers and schedules execution paths based on vector similarity. Ding \textit{et al.}~\cite{ding2024enhancing} store interaction histories for nearest-neighbor retrieval, using subset selection to limit storage overhead. Hybrid-RACA~\cite{xia-etal-2024-hybrid} combines a cloud retriever with a lightweight edge predictor by transmitting compressed memory units for enhanced local inference without continuous cloud access. Xu \textit{et al.}~\cite{xu2025collabrag,li2025eacoragdistributedtieredllm,fatehkia2024traglessonsllmtrenches} introduce an iterative Direct Preference Optimization (DPO) mechanism, where a large model continually guides updates to a small on-device model.

\subsubsection{Retrieval-Augmented Generation with Self-Assessment and Feedback}

\begin{figure}
	\centering
	\includegraphics[width = \linewidth]{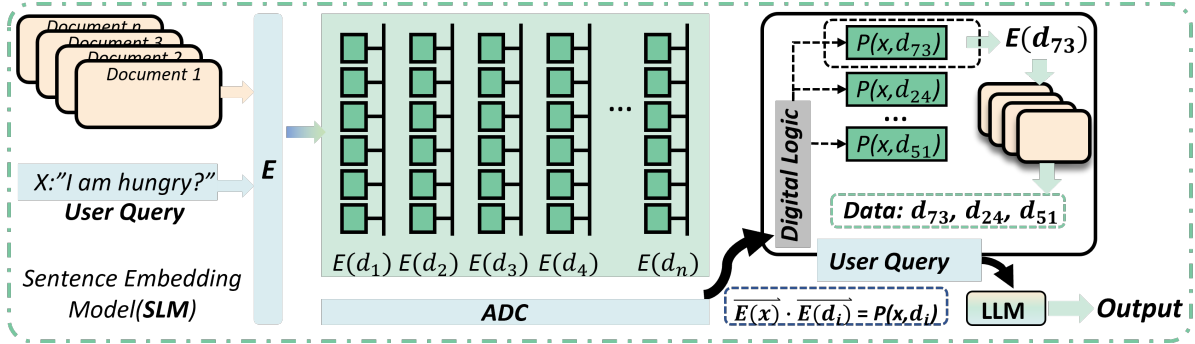}
	\caption{Hybrid SLM-LLM collaboration in RAG: the SLM encodes queries and retrieves documents via approximate nearest-neighbor search, and the LLM generates responses using the query and retrieved context.}
	\label{fig:RAG}
	\vspace{-5pt}
\end{figure}

As shown in \cref{fig:RAG}, these methods leverage external knowledge and reflective reasoning to enhance generation in complex tasks. Qin \textit{et al.}~\cite{10.1145/3736721,harbola2025knowslmframeworkevaluationsmall} analyze trade-offs among fine-tuning, RAG, data scale, model size, and task difficulty, demonstrating that RAG can outperform fine-tuning under resource constraints and that compressed LLMs benefit more from limited personalized data. Self-Knowledge Retrieval~\cite{wang-etal-2023-self-knowledge} prioritizes internal knowledge, invoking external retrieval only when needed via prompt learning or confidence heuristics. Self-RAG~\cite{asai2024selfrag} introduces reflection tokens, special control signals that let the model explicitly assess retrieved passages. CRAG~\cite{yan2024correctiveretrievalaugmentedgeneration} employs a lightweight assessor to evaluate retrieval quality and trigger secondary retrieval or web expansion as necessary. Jiang \textit{et al.}~\cite{jiang2024longragenhancingretrievalaugmentedgeneration} extend introspective signals by allowing the model to either incorporate or recalibrate retrieved content. RA-ISF~\cite{liu-etal-2024-ra} decomposes failed queries into subquestions via a three-stage framework, self-assessment, retrieval, and query decomposition, and integrates subanswers into the final response. SlimRAG~\cite{zhang2025slimragretrievalgraphsentityaware} builds a lightweight entity-to-fragment index on the cloud for salient entity detection, reducing graph complexity and communication. SpeculativeRAG~\cite{wang2025speculativerag} clusters documents for parallel draft generation.

{In practice}, Glocker~\cite{glocker2025llm} applies timestamp-enhanced RAG in autonomous home object management, enabling robots to track past actions while executing high-level instructions. Zhu \textit{et al.}~\cite{10.1145/3676536.3676674} introduce sparse context selection, parallelizing document encoding and decoding only the most relevant cached content via control tokens. SparseRAG~\cite{zhu2025accelerating} integrates document evaluation and response generation to minimize context loading, accelerating inference and improving output quality for both short- and long-form tasks.

\subsection{Mixture: Token-Level Speculation and Verification in Edge-Cloud LLM-SLM Collaboration}
\label{sec:mixture-token}

\begin{figure}
	\centering
	\includegraphics[width = \linewidth]{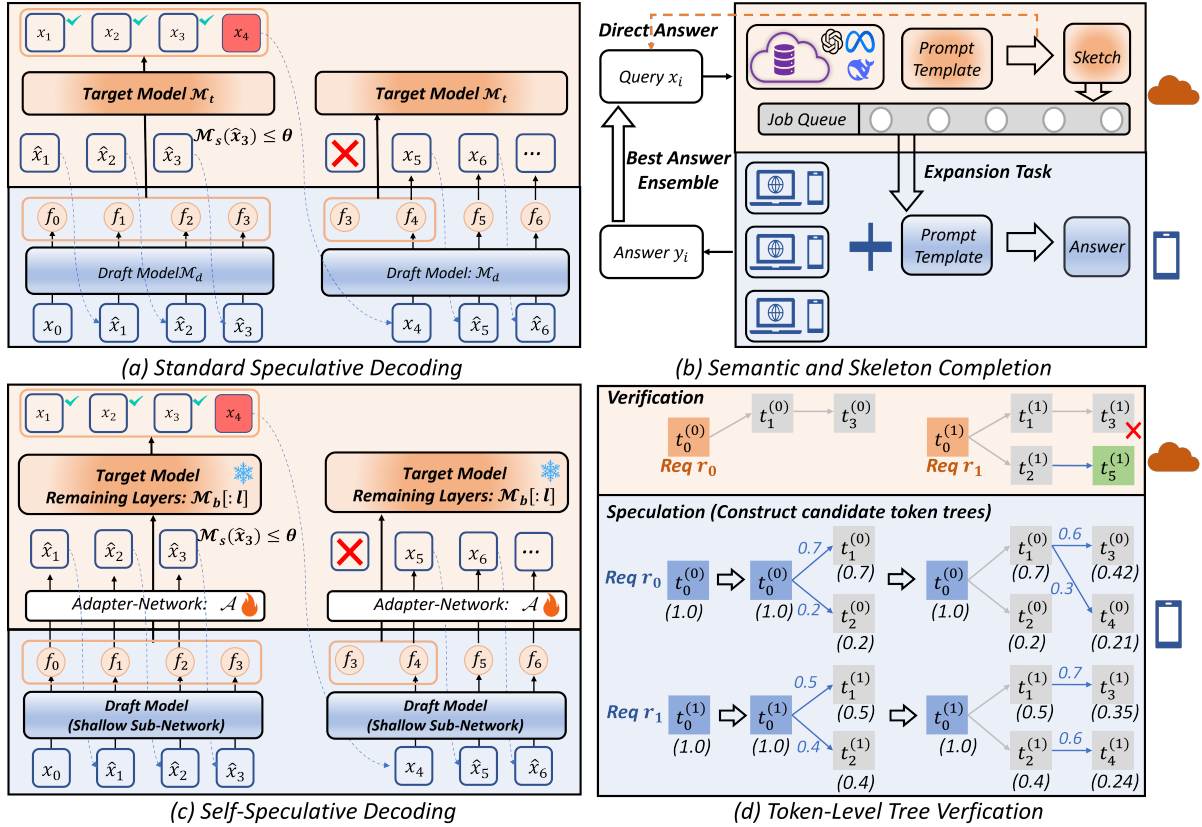}
	\caption{
	{(a) Vanilla:} the cloud verifies edge-generated drafts and rejects or regenerates invalid tokens; 
	{(b) Semantic Skeleton:} the cloud generates high-level semantic skeletons via prompts, which the edge completes with identical or refined prompts; 
	{(c) Self-Speculative:} the model is split into shallow and deep subnetworks, using adapters for lightweight on-device verification; 
	{(d) Token Tree:} token trees are expanded probabilistically, and the cloud verifies optimal branches per request.}
	\label{fig:SD2}
	\vspace{-5pt}
\end{figure}

Most LLMs generate tokens autoregressively, producing each token sequentially for the next prediction. While this yields high-quality output, it incurs substantial latency, which limits real-time deployment on resource-constrained devices. To address this, recent work~\cite{xia-etal-2024-unlocking} proposes a ``lightweight drafting + precise verification'' paradigm: the edge-deployed SLM rapidly generates high-confidence drafts, and the cloud-based LLM verifies semantic consistency, corrects uncertain tokens, and enriches content with additional detail~\cite{zhong2025crossattentionspeculativedecoding}. As shown in \cref{fig:SD2}, we survey speculative decoding architectures and techniques in edge-cloud collaboration, covering draft-refine frameworks, draft-completion strategies, skeleton-completion mechanisms, and token-level verification~\cite{hao2024hybrid}.

\subsubsection{Edge Draft and Cloud Validation}

This line of work divides into two categories: \textit{1)} vanilla speculative decoding frameworks with algorithmic enhancements and \textit{2)} system-level designs that improve parallelism and reduce latency under diverse deployment constraints.

\paragraph{Vanilla Speculative Decoding and Algorithmic Enhancements.} 
Early works such as~\cite{kim2023speculative,kim2025guiding,Gagrani_2024_CVPR} introduced the basic speculative decoding paradigm. A fallback mechanism~\cite{wang2025flashlatentawaresemiautoregressivespeculative} lets the lightweight model defer uncertain tokens to the larger model, while a rollback strategy corrects inaccuracies. Building on this, RSD~\cite{liao2025reward} and SpecExec~\cite{NEURIPS2024_1d91d568} incorporate controlled bias to prioritize high-reward outputs. AutoMix~\cite{aggarwal2024automix} integrates symbolic reasoning with reinforcement learning to mitigate hallucinations~\cite{debortoli2025accelerated} and reasoning errors, using kernel density estimation to close the feedback loop between on-device verification and cloud routing. Fu \textit{et al.}~\cite{fu2025speculative} improve efficiency by alternating the draft and target models as proposer and verifier via a learned verification distribution.

\paragraph{Parallel and Low-Latency Speculative Inference Frameworks.} 
Interactive applications (\textit{e.g.}, question answering, voice assistants, real-time translation) demand low Time-To-First-Token (TTFT) and Token-By-Token Time (TBT). DiSCo~\cite{sun2025disco} estimates the acceptance probability of each speculative token using the drafting model’s logits. To avoid mutual stalling, where the draft model waits for validation, SpecDec~\cite{xia-etal-2023-speculative,10.5555/3618408.3619203} performs parallel token verification with soft discrimination. PEARL~\cite{liu2025pearl} uses adaptive draft lengths and early verification to accelerate decoding, and SEED~\cite{wang-etal-2025-seed} manages drafts with a round-robin FCFS queue.

\paragraph{Multi-Layer and Lightweight Verification Pipelines.} 
To reduce cloud load, \cite{wang2025speculateknowledge,mcdanel2025pipespecbreakingstagedependencies,syu-lee-2025-hierarchical,qian-etal-2024-bass} propose multi-layer pipelines: a lightweight verifier filters token blocks before final cloud validation~\cite{10.5555/3666122.3667436}. Falcon~\cite{DBLP:conf/aaai/GaoXXJ25} boosts intra-block token dependency via coupled sequential scanning, improving speculative accuracy while maintaining a compact two-layer transformer draft model suited for lightweight, multi-layer verification. Spector~\cite{spector2023accelerating} adds an extra draft stage using a simpler model to pre-filter low-confidence hypotheses. ~\cite{anonymous2025fast} uses layer parallelism via early exits for mid-verification token drafting. Clients pre-draft tokens using idle time before final validation, boosting edge-cloud parallelism. This achieves a 21\% speedup over cloud decoding on the Unitree Go 2 robot.

\paragraph{Draft Quality Enhancement and Hardware-Aware Scheduling.} 
Fuzzy Speculative Decoding (FSD)~\cite{holsman2025fuzzy} relaxes strict match constraints, allowing minor mismatches to boost speed~\cite{lin2024efficient}. Zhao \textit{et al.}~\cite{zhao-etal-2024-ouroboros,liu2025hamburger} propose phrase-level reuse, caching validated phrases for rapid generation. DistillSpec~\cite{zhou2024distillspec} distills the target model into the draft model offline using self-sampled data. DuoDecoding~\cite{lv2025duodecoding} assigns drafting to the CPU and validation to the GPU, dynamically adjusting draft lengths based on system load~\cite{10.5555/3692070.3692535}. SPIN~\cite{11044522} uses a learning-based SSM selection without request priors and accelerates execution via GPU pipelining of speculation and verification. BanditSpec~\cite{ BanditSpec} formulated adaptive hyperparameter selection problem when generating text as a multi arm Bandit problem and designed two bandit based hyperparameter selection algorithms.

\subsubsection{Self-Speculative Decoding without Auxiliary Draft Models}

Some studies~\cite{hu2025diffusion,xia2024swift} note that adding a separate draft model increases communication, storage, and training costs~\cite{ou-etal-2024-lossless}. Kangaroo~\cite{NEURIPS2024_16336d94} reuses the target LLM’s shallow subnetwork and LM head for self-drafting, significantly reducing overhead. SWIFT~\cite{xia2024swift} exploits context-aware layer skipping for dynamic computation, and ASD~\cite{hu2025diffusion} extends speculative sampling to diffusion models by leveraging DDPM exchangeability for parallel inference. These methods maintain speed gains without extra model training.

\begin{table}
	\small
	\centering
	\caption{Comparison of cloud-to-edge and edge-to-cloud skeleton collaboration paradigms}
	\begin{tabularx}{\linewidth}{c|l|X|X|X}
	\toprule
	\textbf{Paradigm} & \textbf{Ref.} & \textbf{Key Ideas} & \textbf{Advantages} & \textbf{Limitations} \\
	\midrule
	\multirow{11}[4]{*}{\rotatebox{90}{Cloud-to-Edge}}
	& \multirow{3}{*}{PICE~\cite{zhan2025pice}}
	& Progressive inference: cloud LLM drafts, edge SLM refines 
	& Reduces latency and supports incremental, real-time reasoning 
	& Requires fine-grained synchronization between LLM and SLM \\
	\cmidrule(lr){2-5}
	& \multirow{4}{*}{CoGenesis~\cite{zhang-etal-2024-cogenesis}} 
	& Sketch planning by LLM with local completion; distribution-only collaboration via logits 
	& Enables privacy-preserving inference and adapts to bandwidth constraints 
	& Coarse plans may limit expressiveness; logit protocols are model-specific \\
	\cmidrule(lr){2-5}
	& \multirow{4}{*}{NEST~\cite{li2024nearest}} 
	& Cloud performs full retrieval, edge conducts token-level neighbor search for caching 
	& Reduces corpus storage and improves retrieval latency 
	& Requires careful tuning of retrieval-token matching; introduces multi-stage complexity \\
	\midrule
	\multirow{9}[4]{*}{\rotatebox{90}{Edge-to-Cloud}}
	& \multirow{3}{*}{SlimPLM~\cite{tan-etal-2024-small}} 
	& Triggers LLM only when edge model uncertainty is high 
	& Minimizes cloud calls and provides fast edge responses 
	& Depends on accurate uncertainty detection; may miss subtle errors \\
	\cmidrule(lr){2-5}
	& \multirow{3}{*}{Hao \textit{et al.}~\cite{hao2024hybrid}} 
	& Token correction: LLM refines output by editing a few edge tokens 
	& Efficient corrections with minimal cloud usage 
	& Requires reliable error localization; not suited for structural rewrites \\
	\cmidrule(lr){2-5}
	& \multirow{3}{*}{Probe Sampling~\cite{10.5555/3737916.3739617}} 
	& Measures draft-target similarity to decide cloud fallback 
	& Lowers cloud overhead while maintaining output fidelity 
	& Relies on precise similarity estimation; risk of under-correction \\
	\bottomrule
	\end{tabularx}
	\label{tab:skeleton}
\end{table}

\subsubsection{Semantic and Skeleton Completion}

The draft-refine paradigm splits generation into two stages: edge-side drafting and cloud-side refinement (see \cref{tab:skeleton}).

\paragraph{Cloud-to-Edge Skeletons.} 
This class of methods adopts a top-down strategy, where the cloud LLM first produces a high-level semantic ``skeleton'', and the edge-deployed SLM complements or adapts the content based on local context or constraints. PICE~\cite{zhan2025pice} generates a task sketch for concurrent edge refinement, reducing latency. CoGenesis~\cite{zhang-etal-2024-cogenesis} offers \textit{1)} sketch-based planning with local completion and \textit{2)} logit-based privacy-preserving inference. NEST~\cite{li2024nearest} uses hybrid retrieval in the cloud and token-level neighbor search on the edge via lightweight caches.

\paragraph{Edge-to-Cloud Drafting.} 
This paradigm starts with fast, low-cost draft generation at the edge, followed by refinement from a more capable cloud model. SlimPLM~\cite{tan-etal-2024-small} triggers LLM engagement based on local knowledge sufficiency. Hao \textit{et al.}~\cite{hao2024hybrid} perform token-level corrections, selectively replacing ambiguous tokens. Probe Sampling~\cite{10.5555/3737916.3739617} filters prompts based on draft-target similarity, minimizing cloud usage. While collaboration granularity varies, from document-chunk retrieval to token-level correction, these methods exemplify two complementary coordination paradigms: cloud-to-edge sketching for semantic decomposition and edge-to-cloud refinement for lightweight, responsive drafting.

\subsubsection{Token Tree Verification}
In speculative decoding, optimal verification timing is crucial: early verification may waste computation, while delayed checks risk irreversible errors (see \cref{fig:SD2}). To mitigate this, the token tree structure allows each node to branch into multiple candidate paths, enabling broader output exploration and reducing waste from single-path errors. Along this line, LLMCad~\cite{xu2023llmcad} and Traversal Verification~\cite{weng2025traversalverificationspeculativetree} introduce a non-autoregressive token-tree verifier capable of simultaneously validating and correcting all branches within a single iteration. Building on this foundation, AdaServe~\cite{li2025adaserve} constructs a candidate token tree for each request and dynamically selects the optimal token branch to maximize system throughput under predefined service level objectives (SLOs)~\cite{yang2025longspec,10.1145/3620666.3651335}. OPT-Tree~\cite{wang2025opt} addresses the inefficiency of fixed-structure token trees by proposing a dynamic tree construction method. It greedily builds an expectation-optimal candidate tree at each decoding step and prunes branches via probabilistic modeling, thus maximizing valid tokens under a node budget. Traversal Verification~\cite{weng2025traversalverificationspeculativetree} further enhances tree utilization by introducing bottom-up, sequence-level verification. This method evaluates complete token paths rather than individual tokens, reducing the risk of prematurely discarding useful subsequences. Sequoia~\cite{NEURIPS2024_ea1f5f08} builds on this by formulating token tree construction as a dynamic programming problem. It avoids draft model resampling and integrates hardware-aware optimizations to adaptively select tree depth and size for target platforms. 

\begin{table}
	\small
	\centering
	\caption{Comparison of collaborative training paradigms in edge-cloud LLM-SLM systems}
	\label{tab:training-paradigms}
	\begin{tabularx}{\linewidth}{X|X|X|X}
	\toprule
	\multirow{2}{*}{\textbf{Category}} & \textbf{Definition \& Characteristics} & \multirow{2}{*}{\textbf{Advantages}} & \multirow{2}{*}{\textbf{Limitations}} \\
	\midrule
	Distillation-Based Collaboration 
	& Uses logits, features, or hidden states for LLM-to-SLM supervision with task- and domain-adaptive schemes 
	& Flexible; enables effective model compression and personalization 
	& Sensitive to domain shift and teacher-student mismatch; data quality-dependent; unidirectional by default \\
	\midrule
	Multi-SLM Parameter Fusion 
	& Integrates heterogeneous SLMs via stitching, subnet assembly and capacity-guided transfer 
	& Enhances edge personalization; enables cross-domain parameter reuse; supports one-pass deployment 
	& Structural mismatch sensitivity; requires functional/meta-guidance; faces stability challenges \\
	\midrule
	Adapter-Based Modular Training 
	& Inserts LoRA-like adapters for parameter-efficient fine-tuning and federated updates 
	& Scalable across tasks and devices; reduces communication costs; preserves base model integrity 
	& Depends on shared adapter architectures; limited in cross-modal or non-aligned scenarios \\
	\midrule
	SLM-Driven LLM Supervision 
	& Employs SLM CoT to refine LLM outputs in noisy/low-resource contexts 
	& Improves interpretability; enhances reliability on simple or repetitive tasks 
	& Constrained by SLM capacity; risk of reinforcing local biases \\
	\midrule
	Cloud-Guided Capability Injection 
	& LLMs generate task-specific modules or behavioral updates for SLMs via meta-learning or distillation 
	& Enables few-shot adaptation; supports evolving edge demands; modular and task-aware 
	& Requires costly cloud computation; may induce concept drift; on-device validation is challenging \\
	\bottomrule
	\end{tabularx}
\end{table}

\section{Overview of Collaborative learning Architectures}
\label{Collaborative-learning-Architectures}

Collaborative training facilitates ongoing knowledge exchange between edge SLMs and cloud LLMs under heterogeneous conditions. Unlike latency-focused inference cooperation, training stresses structural compatibility, non-IID adaptation, and privacy-preserving transfer. Traditional distillation and transfer learning assume aligned architectures or tasks, limiting generality. In practice, structural gaps and data heterogeneity cause biases and spurious correlations that harm aggregation and generalization. Liu \textit{et al.}~\cite{liu2025harness} formalize large-small model collaboration under privacy, exposure, and cost constraints, decomposing it into three stages: \textit{1)} LLM-to-SLM transfer, \textit{2)} SLM-to-LLM feedback, and \textit{3)} closed-loop adaptation. To unify this landscape, we propose a taxonomy of collaborative training across five paradigms: distillation-based collaboration, adapter-based modular training, bidirectional learning, SLM-driven supervision, and cloud-guided capability injection. As shown in \cref{tab:training-paradigms}, our classification highlights each paradigm’s functional role and progressive interdependence, from unidirectional transfer to modular tuning, iterative feedback, and reversed supervision. Grounded in knowledge flow, coupling strength, structural assumptions, and deployment constraints, this taxonomy provides a principled basis for selecting suitable strategies in edge-cloud contexts and inspires more robust, generalizable training designs.

\begin{figure}
	\centering
	\includegraphics[width = \linewidth]{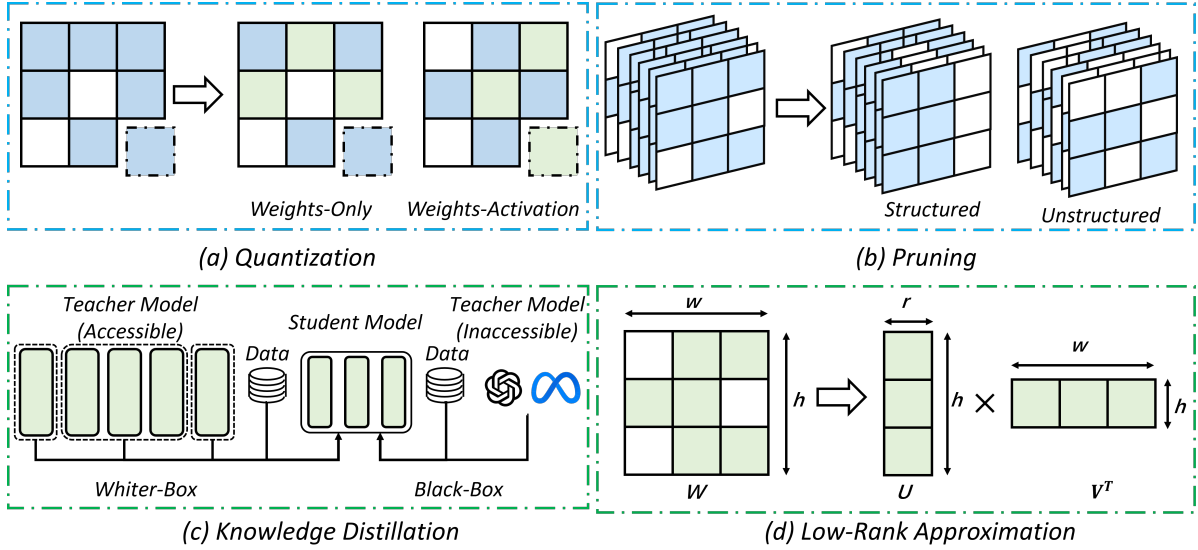}
	\caption{Model compression techniques for efficient cloud-edge collaborative training: (a) quantization of weights and activations; (b) pruning of parameters (structured or unstructured sparsity); (c) knowledge distillation for teacher-student transfer; and (d) low-rank approximation of weight matrices.}
	\label{fig:KD-Pruning-quantization}%
	\vspace{-5pt}
\end{figure}

\subsection{Pruning and Quantization Strategies for Efficient Device-Cloud Collaboration}

In device-cloud collaborative training frameworks, pruning and quantization are key techniques for constructing lightweight models that reconcile edge devices’ limited resources with cloud models’ accuracy demands. Li \textit{et al.}~\cite{li2023cloud} introduce a sparsity-aware channel pruning method that evaluates feature distribution deviations to remove globally unimportant channels; its soft-mask mechanism allows selective reactivation. EfficientLLM~\cite{xing2025efficientllmscalable} applies progressive structural pruning during pretraining to eliminate redundant substructures. Complementing pruning, Jiang \textit{et al.}~\cite{jiang2023high} propose a split-transformer design in which a lightweight edge encoder produces quantized intermediate embeddings for cloud decoding. MergeNet~\cite{li2025mergenet} addresses structural incompatibility via low-rank decomposition of LLM and SLM parameters, followed by attention-based fusion, enabling expressive knowledge transfer without altering the edge architecture. As illustrated in \cref{fig:KD-Pruning-quantization}, these methods co-optimize pruning and quantization across training, inference, and scheduling dimensions. For task-granularity mixtures in edge-cloud inference, the OT-GAH algorithm~\cite{10759588} leverages online tree pruning and generalized assignment heuristics to achieve near-optimal throughput under diverse latency and accuracy constraints.

\subsection{Distillation and Low-Rank Approximation}

Edge-cloud distillation techniques fall into three broad categories: \textit{1)} task- and domain-adaptive distillation, \textit{2)} architectural decoupling with proxy learning, and \textit{3)} bidirectional, feedback-driven distillation.

\paragraph{Task- and Domain-Adaptive Distillation.} 
ATKD redefines distillation along task- and diversity-oriented axes~\cite{zhong-etal-2024-revisiting}, introducing an uncertainty coefficient to quantify token-level learning difficulty and revealing that high LLM output certainty suppresses diversity. Methods such as SLMREC~\cite{xu2025slmrec} and GKT~\cite{yao2024gkt} align intermediate hidden states between lightweight on-device SLMs and cloud LLMs for latent-space transfer. Domain-robust techniques~\cite{DBLP:conf/iclr/TangL0ZD0K24,gu2023knowledge} construct a pseudo-sample space, using latent mapping, masking, and Mixup, to align student representations without shared data. DDK~\cite{NEURIPS2024_b206d54f} further improves cross-domain robustness via domain-guided sampling and factor-smoothing mechanism to facilitate efficient cloud-to-edge distillation across diverse domains.

\paragraph{Architectural Decoupling and Feedback-Guided Distillation.} 
To decouple deployment between edge and cloud, DC-CCL~\cite{ding2023dc} vertically partitions a base model into cloud and edge components, training a lightweight proxy via distillation to mimic cloud behavior with minimal communication. Starodubcev \textit{et al.}~\cite{10655604} adaptively invoke teacher corrections only when student outputs fail quality checks. Collin \textit{et al.}~\cite{10.5555/3692070.3692266} integrate weak supervision and high-capacity teachers into a closed-loop framework, while Co-Supervised Learning~\cite{liu2024cosupervise} dynamically allocates teachers and filters noisy signals via posterior consistency. Inverted supervision methods like SALT~\cite{rawat2024littlehelp} and SKD~\cite{xu2024speculative} have edge SLMs teach LLMs early in training, gradually reversing roles to focus LLMs on complex samples. Multi-modal, feedback-enhanced pipelines, CD-CCA~\cite{wang2024cloud} and LLM-QAT~\cite{liu-etal-2024-llm}, upload utility-selected samples for cloud optimization and return quantized updates to the edge, enabling efficient, compressible, and adaptive model enhancement. Collectively, these methods demonstrate a progression from static partitioning to dynamic, bidirectional, and multi-modal knowledge transfer in edge-cloud collaborative training.

\paragraph{Leveraging Low-Rank Approximation for Edge-side SLMs Deployment}. Beyond knowledge distillation, low-rank approximation has emerged as a lightweight and effective approach for optimizing LLM deployment in resource-constrained edge environments. QLLMS~\cite{DBLP:conf/infocom/HuHW25} addresses edge performance unpredictability by reconstructing the AQS matrix from partial samples via a low-rank attribute-driven recovery method. DP-LoRA~\cite{DBLP:journals/tmis/LiuZZGZWQ25} reduces transmission overhead in distributed training through low-rank weight updates. Moreover,~\cite{11020604} enhances stability-aware fine-tuning using fractional programming and an iterative-level penalty (IRP) to jointly optimize resource allocation and user–edge association.

\subsection{Parameter Compatibility and Model Convergence}

Traditional unified model-transfer methods struggle with parameter mismatches and personalized edge requirements. To address fragmented domain knowledge in collaborative learning, recent studies propose efficient parameter-coordination mechanisms:
Graft~\cite{dai2025graftintegratingdomainknowledge} introduces a compatibility-aware stitching strategy, using local functional attribution and global information-theoretic signals to integrate only compatible parameters. CKI~\cite{Lv_Ye} evaluates a source model’s information capacity and fuses compatible parameters into the target via a two-stage transfer and stitching process.
Forward-OFA~\cite{10.1145/3690624.3709178} uses real-time edge demands to assemble sub-networks through behavior-to-structure mapping, resolving gradient conflicts and allowing one-pass adaptation without backpropagation. DIET~\cite{10.1145/3637528.3671669} maintains a unified backbone across devices, while the cloud generates personalized ``diets'' (subnetworks) based on each device’s usage history.
FedMKT~\cite{fan-etal-2025-fedmkt} explores two-way knowledge exchange between cloud LLMs and client SLMs, improving adaptation under non-IID data. ModelGPT~\cite{tang2024modelgpt} leverages LLMs to automatically generate customized small models from user-provided samples or task descriptions.

\subsection{Adapter-Based Modular Designs}

In collaborative edge-cloud systems, integrating knowledge and weight parameters between SLMs and LLMs remains a core challenge. To address this, adapter methods (\textit{e.g.}, LoRA) enable scalable, parameter-efficient tuning across heterogeneous edge-cloud systems:
PEFT~\cite{fan2024fedcollmparameterefficientfederatedcotuning} uses lightweight LoRA modules as communication bridges: clients update only adapter weights during federated training, while the server performs bidirectional knowledge extraction via supervised fine-tuning and KL-divergence regularization.
Lu \textit{et al.}~\cite{lu-etal-2021-parameter-efficient} insert bottleneck adapters between Transformer layers to encode domain knowledge into pre-trained models. PLURALISM~\cite{feng-etal-2024-modular} integrates community LLMs into a foundation model via LoRA, enabling edge SLMs to fine-tune locally and the cloud LLM to manage integration.
HETLoRA~\cite{cho-etal-2024-heterogeneous} combines high- and low-rank LoRA modules; clients rank modules by capability, and the server applies rank-aware pruning and sparsity-weighted aggregation. CDC-MMPG~\cite{10.1145/3706422} introduces a Fast Domain Adapter that uses historical multi-modal data to train a global model and generates personalized parameters for real-time device inputs.
Collectively, these approaches mark a shift toward flexible, modular, and data-aware tuning mechanisms that seamlessly integrate knowledge across end-cloud collaborative frameworks.

\begin{figure}
	\centering
	\includegraphics[width = \linewidth]{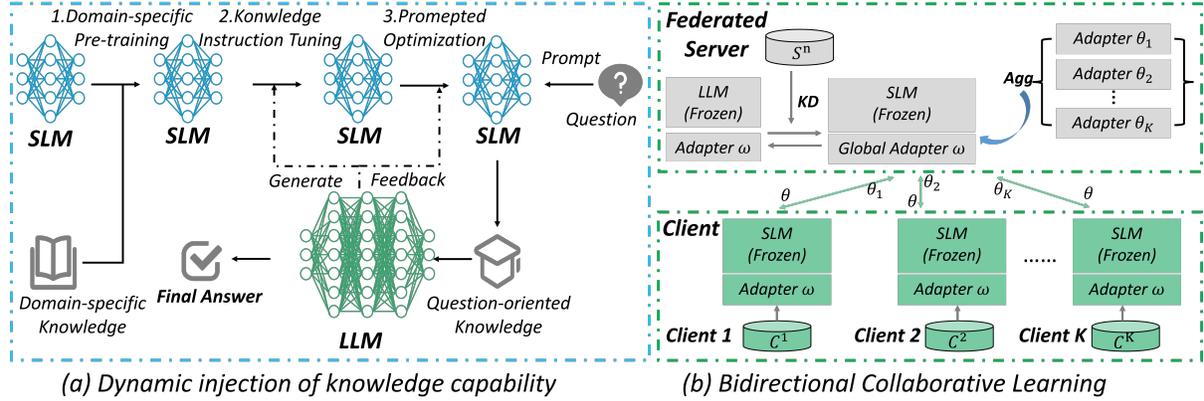}
	\caption{Collaborative training with bidirectional knowledge transfer: (a) the SLM acquires task knowledge through domain-specific pretraining, instruction tuning, and prompt optimization; (b) frozen client SLMs augmented with local adapters train collaboratively under centralized coordination.}
	\label{fig:knowledge-transfer}%
	\vspace{-5pt}
\end{figure}

\subsection{Bidirectional Collaborative Learning with Heterogeneous Models}

Most distribution-shift methods rely on on-device personalization via online training or gradient updates, which incur high latency and overhead. To address this, recent approaches favors collaborative optimization across heterogeneous edge-cloud models, emphasizing iterative, bidirectional knowledge transfer over static, one-way distillation (see \cref{fig:knowledge-transfer}):
DUET~\cite{10.1145/3543507.3583451} separates static and dynamic layers, using a Universal Meta-Network and hierarchical HyperNetworks to generate adaptive parameters for personalized edge inference.
Hu \textit{et al.}~\cite{hu2024dynamicensemblereasoningllm} develop a knowledge-transfer prompting technique that lets different LLMs share and integrate complementary knowledge effectively.
CROSSLM~\cite{deng2023mutual} establishes a pipeline where edge models train locally, and the cloud generates and filters pseudo-data using edge feedback, enabling mutual enhancement.
SLM~\cite{lin2025efficientmulti} applies inverse reinforcement learning on the edge to produce structured, task-aligned samples, while cloud LLMs perform parallel-decoding distillation to align global and local objectives.

\subsubsection{Dynamic Knowledge Transfer from Cloud LLMs to On-Device SLMs}

This research line treats small models as pluggable knowledge modules that supplement or enhance LLM performance during inference. Knowledge Card~\cite{feng2023knowledge} uses a pretrained SLM to generate ``knowledge cards'', a parameterized knowledge base dynamically queried during inference. BLADE~\cite{li2025blade} employs a plugin-style architecture where domain-specific SLMs capture expert knowledge and fuse it with general-purpose LLMs via Bayesian optimization to improve reasoning quality. Progressive Distillation~\cite{hsieh-etal-2023-distilling} targets unlabeled data by having cloud LLMs generate task labels and rationales to train lightweight edge models, jointly optimizing label and rationale losses for improved accuracy and interpretability. The HEF framework~\cite{yang2024enhancingempatheticresponsegeneration} integrates a small-scale Empathy Model to detect emotions and triggers, guiding LLM generation for more nuanced dialogue. FedCFA~\cite{jiang2025fedcfa} addresses non-IID generalization in federated settings by constructing a global feature dataset and applying causality-aware counterfactual replacement, aligning SLMs with cloud semantics. AcKnowledge~\cite{das-etal-2024-acknowledge} leverages user-question-driven meta-learning to acquire external knowledge and continually refine SLMs with user feedback when distortions are detected.

\subsubsection{Adaptive Capability Injection into Small Models via Cloud Guidance}

This category reverses the typical distillation flow, using SLMs as ``teachers'' or behavioral priors to guide LLM training and fine-tuning. ECLM~\cite{zhuang2023eclmefficientedgecloudcollaborative} decomposes cloud models into composable modules, assembling task-specific submodels for diverse devices and periodically integrating new edge knowledge back into the cloud. Mitchell \textit{et al.}~\cite{mitchell2023emulator} encode fine-tuning-induced behavioral changes as a compact ``logit delta'', which a lightweight SLM computes and transfers to an LLM to emulate these adaptations without full retraining. Tang \textit{et al.}~\cite{tang2024small,10.5555/3737916.3738072} present an entity-relation extraction system where a small PLM captures head-class knowledge and generates intermediate predictions serving as chain-of-thought guidance for the LLM. Purify-LLM~\cite{li2024purifyinglargelanguagemodels} filters noisy data for LLMs using a trusted SLM and the CP-$\delta$ algorithm, integrating only aligned knowledge. These methods highlight SLMs’ role in constraining and guiding LLM learning, creating a positive feedback loop that optimizes model behavior along explicit paths.

\section{Benchmarks, Datasets, and Evaluation Protocols}
\label{Benchmarks-Datasets-and-Evaluation-Protocols}

Evaluating collaborative LLM-SLM systems is more complex than centralized models due to a lack of standardized benchmarks for edge-cloud collaboration. Traditional NLP benchmarks assume IID data and centralized execution, unlike real deployments where SLMs manage personalized, non-IID workloads. Thus, new benchmarks must enable user- or device-level partitioning to reflect deployment heterogeneity.

\begin{table}
	\small
	\centering
	\caption{Representative datasets for edge-cloud collaboration}
	\label{tab:Learning2}
	\begin{tabularx}{\linewidth}{l|X|X|X|X}
	\toprule
	\textbf{Dataset} & \textbf{Task} & \textbf{Volume \& Classes} & \textbf{Partitioning} & \textbf{Metrics \& Features} \\
	\midrule
	\multirow{3}{*}{LEAF~\cite{caldas2019leafbenchmarkfederatedsettings}}
	& Multi-task federated learning 
	& Sent140: 660k tweets; FEMNIST: 805k images 
	& User-level splits reflecting real-world heterogeneity 
	& Supports personalization and transfer learning \\
	\midrule
	\multirow{3}{*}{iNaturalist-User-120k~\cite{10.1007/978-3-030-58607-2_5}}
	& Image classification 
	& 120,300 images from 9,275 users; 1,203 species 
	& Uploader-ID splits (user-level) 
	& Large label space with strong user-level statistical skew \\
	\midrule
	\multirow{3}{*}{Landmarks-User-160k~\cite{10.1007/978-3-030-58607-2_5}} 
	& Landmark recognition 
		& 164,172 images from 1,262 photographers; 2,028 landmarks 
	& Region-aware S2-cell splits via GPS metadata 
	& Captures spatial distribution heterogeneity \\
	\midrule
	\multirow{3}{*}{PersonalDialog~\cite{zheng2019personalized}} 
	& Personalized dialogue 
	& 56.3 M utterances from 8.5 M speakers 
	& User-ID splits with demographic annotations 
	& Supports sociolinguistic and attribute personalization \\
	\midrule
	\multirow{3}{*}{PERSONA-CHAT~\cite{zhang-etal-2018-personalizing}} 
	& Persona-grounded chit-chat 
	& 164,356 utterances across 10,981 dialogues 
	& Crowdworker pairs on 1,155 personas 
	& Next-utterance prediction and personal consistency \\
	\midrule
	\multirow{3}{*}{LiveChat~\cite{gao-etal-2023-livechat}} 
	& Multi-party live dialogue 
	& 1.33 M utterances from 351 streamers 
	& ASR-aligned utterances with recipient matching 
	& Single-turn dialogue with personas; addressee recognition \\
	\midrule
	\multirow{3}{*}{FedScale~\cite{10.1145/3477114.3488760}} 
	& Multi-domain federated learning 
	& 20 datasets (CV, NLP, audio) with millions of clients 
	& Non-IID splits by user ID or trace logs 
	& Benchmarks statistical; System heterogeneous, dynamics \\
	\midrule
	\multirow{3}{*}{FedNLP~\cite{lin-etal-2022-fednlp}} 
	& Federated NLP (TC, NER, QA, Seq2Seq) 
	& 20News: 11.3k, 20 classes; OntoNotes: 50k, 37 tags
	& Dirichlet non-IID splits and cross-silo partitions 
	& Metrics: accuracy, F1, ROUGE; federated evaluation \\
	\bottomrule
	\end{tabularx}
\end{table}

\subsection{Datasets for Federated Edge-Cloud Collaboration}

Realistic, heterogeneous, and reproducible benchmarks are crucial for evaluating collaborative training and inference in edge-cloud LLM-SLM systems (see \cref{tab:Learning2}). The LEAF framework~\cite{caldas2019leafbenchmarkfederatedsettings} provides modular tools and real-world, user-partitioned datasets to assess performance under statistical heterogeneity, resource constraints, and personalization. For large-scale visual tasks, Zhu \textit{et al.}~\cite{10.1007/978-3-030-58607-2_5} introduce iNaturalist-User-120k (120K images, 1.2K species, 9.3K users) and Landmarks-User-160k (164K images, 2K categories, 1.3K uploaders) with non-IID partitions. In dialogue modeling, PersonalDialog~\cite{zheng2019personalized} offers 56 M utterances from 8.5 M speakers annotated with demographic attributes, while PERSONA-CHAT~\cite{zhang-etal-2018-personalizing} provides persona-based consistency benchmarks. LiveChat~\cite{gao-etal-2023-livechat} supplies 1.33 M multi-party Chinese dialogues with 351 roles, enabling response generation and recipient identification under temporal dynamics. FedScale~\cite{10.1145/3477114.3488760} covers 18 real-world tasks across modalities and scales from hundreds to millions of clients, with its FAR platform supporting asynchronous training, latency simulation, and metrics for accuracy, cost, and privacy. FedNLP~\cite{lin-etal-2022-fednlp} benchmarks federated NLP tasks (\textit{e.g.}, 20Newsgroup, MRQA) using Dirichlet splits to simulate non-IID distributions, facilitating reproducible evaluation and real-world optimization of collaborative LLM-SLM systems.

\subsection{Benchmarks for Dual-Model Edge-Cloud Learning}
\label{Benchmark for Edge-Cloud Collaborative Learning with Dual-Model Architecture}

Edge-cloud collaborative learning demands dual evaluation: global generalization by the cloud LLM and local adaptation by edge SLMs, under non-IID data across devices. Recent benchmarks create non-IID test environments via user-level splits. LEAF extends CIFAR-10 and Shakespeare into federated versions with device-ID partitions~\cite{caldas2019leafbenchmarkfederatedsettings}. Persona-Chat~\cite{zhang-etal-2018-personalizing} and LiveChat~\cite{gao-etal-2023-livechat} exploit dialogue roles for personalized generation. FedMulti-modal~\cite{10.1145/3580305.3599825} integrates ten datasets across eight modalities, simulating missing modalities and label noise. FederatedScope-GNN~\cite{10.1145/3534678.3539112} offers heterogeneous graph benchmarks (\textit{e.g.}, FedDBLP) with node-, edge-, and graph-level splits. pFL-Bench~\cite{NEURIPS2022_3cc03e19} unifies partition strategies across 10+ datasets and introduces device heterogeneity and sparsity configurations for realism.

Hierarchical evaluation metrics are vital: image classification employs weighted global accuracy (by device data volume), local SLM accuracy, and distillation loss to assess knowledge transfer~\cite{bonawitz2019federatedlearningscaledesign}. For NLG, global BLEU/ROUGE scores complement user-level perplexity. In recommendations, global AUC weighted by user data and local CTR correction reflect personalization~\cite{ding2022device}, with OCPC evaluating end-to-end traffic-allocation efficiency~\cite{10.1145/3097983.3098134}. 

Open-source platforms support reproducible edge-cloud experiments: FedML~\cite{he2020fedmlresearchlibrarybenchmark} offers on-device, single-machine, and distributed modes with modular communication-training decoupling; Flower~\cite{beutel2022flowerfriendlyfederatedlearning} scales to millions of virtual clients across languages and backends; Google’s TensorFlow Federated~\cite{bonawitz2019federatedlearningscaledesign} runs on hundreds of millions of devices. For LLM-specific operations, LLMOps frameworks~\cite{info16020087,DBLP:journals/corr/abs-2402-05333} introduce tailored monitoring and safety protocols. SpecBench~\cite{xia-etal-2024-unlocking} evaluates speculative decoding across latency, accuracy, and compute cost. MessageRewriteVal~\cite{zhu-etal-2024-towards} standardizes mobile text rewriting with high-quality, human-annotated scenarios to assess a model’s ability to rewrite messages based on natural language instructions. Together, these benchmarks and platforms establish a foundation for fair, rigorous evaluation of collaborative LLM-SLM systems under realistic edge-cloud conditions.

\section{Open Challenges}
\label{Open Challenges}

\subsection{Privacy-Preserving and Secure Collaboration Mechanisms}

Traditional optimizations for inference efficiency often overlook privacy and security risks in collaborative training and updating~\cite{10.1145/3708528}. Achieving efficient edge-cloud cooperation while preserving data locality has thus become a critical research direction, with secure multi-party computation encrypting gradient updates to prevent sensitive information leakage during aggregation.

\paragraph{Training-level protections.} 
Liu \textit{et al.}~\cite{liugrey} integrate edge-side prompt adaptation with cloud-based optimization via lightweight adapters, enabling privacy-preserving collaboration by keeping sensitive data local while leveraging cloud resources for efficiency. DCPR~\cite{long2024diffusion} adopts a hierarchical diffusion paradigm for personalized recommendation, where general patterns are learned in the cloud, regional preferences are adapted at the edge, and fine-grained personalization occurs on-device. Luo \textit{et al.}~\cite{luo2025federatedlearningbaseddatacollaboration} utilize federated learning to train models across decentralized clients, ensuring privacy by transmitting only encrypted updates. Building upon these advances, future efforts will increasingly center on uncertainty-guided supervision for selective cloud involvement, fair aggregation under skewed and non-\textit{i.i.d.} client distributions, and generalizable knowledge transfer across heterogeneous tasks, modalities, and model scales. Furthermore, counterfactual representation learning is expected to play a pivotal role in debiasing cloud-side updates before aggregation, enhancing both utility and fairness in collaborative training.

\paragraph{Inference-level protections.} 
RemoteRAG~\cite{cheng2024remoteragprivacypreservingllmcloud} introduces a semantic differential-privacy metric in embedding space and a dynamic secure retrieval protocol that only returns document indices when similarity exceeds a threshold. SuperICL~\cite{xu2023small} uses plugin-style small models to inject local knowledge via prompt augmentation, guiding the LLM’s reasoning without exposing raw data. Pan \textit{et al.}~\cite{pan2024cloud} propose a hybrid framework where the cloud builds proxy models with adapters and compression, and the edge applies symbolic masking to sensitive content, generating substitute data via cGANs to balance utility and privacy. POST~\cite{li2025efficient} accelerates privacy-preserving speculative decoding by offloading draft generation to a public GPT model and optimizing cryptographic operations. Chen \textit{et al.}~\cite{10707974} orchestrate a hybrid Kubernetes-based system that retains sensitive data and vector stores (\textit{e.g.}, Chroma) in private clouds while elastically offloading LLM inference to public clouds under strict confidentiality constraints. These efforts provide initial approaches to privacy-preserving collaboration but reveal open challenges. Future work should scale secure, efficient speculative decoding across diverse model hierarchies, balance draft quality and verification cost in multi-model setups, ensure robustness amid noisy feedback and unstable connections, and enable emergent capabilities through self-reflective SLMs within larger collaborative systems.

\begin{table}
	\small
	\centering
	\caption{Industrial cloud-edge LLM-SLM collaboration frameworks and application scenarios}
	\label{tab:application}
	\begin{tabularx}{\linewidth}{c|l|X|X}
	\toprule
	\textbf{Category} & \textbf{Domain} & \textbf{Collaboration Strategy} & \textbf{Advantages \& Limitations} \\
	\midrule

	\multirow{9}[4]{*}{\rotatebox{90}{\makecell{Industrial Edge-Cloud \\ Frameworks}}}
	& \multirow{3}{*}{Walle~\cite{lv2022walle}} 
	& End-to-end deployment pipeline handling development, runtime, and scaling. 
	& Manages 300+ tasks and 10B+ daily calls; constrained by tight infrastructure coupling. \\
	\cmidrule(lr){2-4}
	& \multirow{3}{*}{Luoxi~\cite{luoxi2024models}} 
	& Slow-fast learning: cloud LLM generates aux. reps; edge SLM fast inference with feedback. 
	& Enhances personalization via feedback loop; requires robust bidirectional communication. \\
	\cmidrule(lr){2-4}
	& \multirow{3}{*}{InfiGUIAgent~\cite{liu2025infiguiagentmulti-modalgeneralistgui}} 
	& Two-stage hierarchical reasoning with on-device fine-tuning for multi-modal GUI interaction. 
	& Enables on-device GUI reasoning; may face edge memory and compute limitations. \\

	\midrule

	\multirow{26}[16]{*}{\rotatebox{90}{Vertical Applications}}
	& \multirow{3}{*}{Autonomous driving~\cite{hu2024cloud,chen2024edge}} 
	& Edge SLM handles perception; cloud LLM performs high-level planning and 	reasoning. 
	& Balances fast perception with complex strategy; requires efficient cloud triggering. \\
	\cmidrule(lr){2-4}
	& \multirow{3}{*}{Livestream product recognition~\cite{10415802}} 
	& Edge extracts keyframe features; cloud LLM conducts multi-modal classification. 
	& Reduces bandwidth via selective uploads; risks missing critical content in keyframes. \\
	\cmidrule(lr){2-4}
	& \multirow{3}{*}{Cultural heritage restoration~\cite{zhang2024llmco4mr}} 
	& Edge proposes fragment matches; cloud LLM ranks and aligns results. 
	& Supports time-sensitive, high-accuracy restoration; depends on fragment proposal quality. \\
	\cmidrule(lr){2-4}
	& \multirow{3}{*}{Product modeling\cite{xu2024mpod123}}~ 
	& Edge injects geometric priors; cloud completes texture synthesis. 
	& Improves partial-view generation; limited by cross-device alignment noise. \\
	\cmidrule(lr){2-4}
	& \multirow{3}{*}{Virtual assistants~\cite{openai2024gpt4technicalreport,glm2024chatglmfamilylargelanguage}} 
	& Edge SLM manages routine interactions; cloud LLM handles complex or fallback 	queries. 
	& Ensures responsiveness; cloud fallback may disrupt conversational continuity. \\
	\cmidrule(lr){2-4}
	& \multirow{3}{*}{Personalized recommendation~\cite{lv2025collaboration,lin2024efficient,long2024diffusion}} 
	& Cloud LLM generates candidate items; edge re-ranks using real-time user context. 
	& Adapts swiftly to user intent; candidate diversity hinges on LLM quality. \\
	\cmidrule(lr){2-4}
	& \multirow{3}{*}{Mobile task automation~\cite{10.1145/3636534.3649379}} 
	& Cloud LLM parses app semantics; edge executes tasks via dynamic analysis. 
	& Enables cross-app automation without manual setup; depends on accurate semantic parsing. \\
	\cmidrule(lr){2-4}
	& \multirow{3}{*}{Healthcare~\cite{labrak-etal-2024-biomistral}} 
	& Biomedical LLM fine-tuned from general LLM for on-premise or hybrid deployment. 
	& Offers domain-specific open-source model; needs careful tuning for sensitive settings. \\
	\cmidrule(lr){2-4}
	& \multirow{2}{*}{Web interaction~\cite{gur2024a}} 
	& Cloud LLM decomposes tasks and generates executable scripts. 
	& Automates multi-step web tasks; relies on precise decomposition. \\
	\bottomrule
	\end{tabularx}
\end{table}

\subsection{Application of Edge-Cloud Large-Small Models in Vertical Domains}

Edge-cloud LLM-SLM collaborations demonstrate strong potential across diverse application domains (see \cref{tab:application}). In autonomous driving, ADAS~\cite{hu2024cloud} uses a lightweight multi-modal model (CogVLM2) at the edge for latency-critical tasks (object detection, obstacle avoidance) while a cloud LLM (ChatGPT-4o) handles high-level route planning. EC-Drive~\cite{chen2024edge} extends this with an event-driven design: the edge LLM manages routine driving, triggering the cloud LLM upon distribution shifts (\textit{e.g.}, novel obstacles) for commonsense reasoning and cross-modal understanding.

Live streaming challenges real-time product recognition with low latency, high concurrency, and costly multi-modal inference. Recent frameworks tackle this by using uni-modal edge models to selectively upload keyframes for cloud multi-modal processing~\cite{10415802}. For example, LLMCO4MR~\cite{zhang2024llmco4mr} treats manuscript restoration as a combinatorial task, using on-device neural solvers for fragment selection and cloud LLMs for confidence scoring and alignment. BoFiCap~\cite{10.1007/978-3-031-44693-1_37} decouples image captioning into boundary detection on-device and non-autoregressive filling in the cloud. VITA-1.5~\cite{fu2025vita} employs a three-stage training pipeline (vision-language pretraining, audio alignment, speech decoding) to avoid modality conflicts, while MPOD123~\cite{xu2024mpod123} injects geometric priors at the edge and offloads complex visual generation to the cloud.

Recommendation systems such as cenarios~\cite{10.1145/3534678.3539123}, LSC4Rec~\cite{lv2025collaboration} use cloud LLMs to generate diverse candidate lists, with lightweight edge ranking models adapting in real time to user preferences. AutoDroid~\cite{10.1145/3636534.3649379} automates mobile workflows by combining LLM commonsense reasoning with dynamic analysis of app interfaces. In healthcare, BioMistral~\cite{labrak-etal-2024-biomistral} specializes a general LLM via continued pretraining on biomedical texts to support domain-specific tasks. For web-based interactions, WebAgent~\cite{gur2024a} addresses open-domain variability and the lack of HTML inductive biases by decomposing user instructions into structured subtasks, summarizing lengthy documents, and generating executable Python scripts to perform complex operations. Together, these vertical applications illustrate the transformative impact of hierarchical edge-cloud LLM-SLM architectures across real-world scenarios.

\section{Conclusion}
\label{Conclusion}

This survey offers the first comprehensive review of edge-cloud collaboration between large and small language models (SLMs), addressing both inference-time cooperation and training-time coordination. We propose a unified taxonomy of inference paradigms, task assignment, task division, and hybrid approaches, further refined into task- and token-level granularities. On the training side, we identify key enablers such as bidirectional knowledge transfer, quantization, pruning, and low-rank adaptation, which support efficient deployment without compromising performance. By grounding our analysis in recent advances, we bridge algorithmic techniques with system-level requirements, providing methodological insights and practical guidance for building scalable, low-latency, resource-aware LLM-SLM systems. This unified perspective establishes a solid foundation for future research and deployment in heterogeneous environments.

\textbf{Future Prospects}. In edge-cloud collaborative inference, many systems rely on uncertainty estimation to support dynamic decisions such as early exit, fallback, or model switching, typically by comparing uncertainty between edge-side SLMs and cloud LLMs. However, mainstream methods based on sampling consistency or token-level probabilities face inherent limitations. Normalized softmax probabilities obscure the raw evidential strength encoded in logits, often failing to distinguish whether a model is confident, uncertain, or unfamiliar with the input. This misalignment is particularly problematic in open-ended language generation, where multiple valid continuations may exist. We advocate a future shift toward evidence-based uncertainty estimation, which leverages unnormalized logits to preserve the model’s accumulated training experience. By decomposing uncertainty into epistemic and aleatoric components, \textit{e.g.}, via Dirichlet-based formulations, edge models can more accurately characterize their own reliability. Such evidence-aware strategies offer a finer-grained basis for trust calibration, response escalation, and offloading in collaborative LLM-SLM deployments, enabling more robust and semantically aligned inference.

Based on evidence-driven uncertainty estimation, future collaborative frameworks will shift from fixed rules to adaptive, intelligent policies. Reinforcement and meta-learning can optimize collaboration using task needs, resource limits, and real-time feedback. Cooperation will deepen from simple task splitting to joint optimization and shared representations, enabling finer coordination. In complex fields like multimodal and embodied intelligence, lightweight edge models will handle real-time sensing while cloud LLMs perform high-level reasoning, balancing latency and decision complexity. Challenges remain in ensuring robust collaboration, reducing communication costs, and enabling personalization. Future systems must balance performance, flexibility, and privacy for truly effective edge-cloud LLM-SLM collaboration.

\bibliographystyle{unsrt}
\bibliography{survey}
\end{document}